\documentclass[fleqn,twoside]{niaauthm}

\usepackage{graphicx}
\usepackage{amsmath}
\usepackage{xspace}
\usepackage{dcolumn}
\newcolumntype{d}{D{.}{.}{-1}}
\usepackage{txfonts}

\newcommand{\E}[1]{$\,\times\,10^{#1}$\xspace}
\newcommand{\MnOx}{M\lowercase{n}O$_{\text{x}}$\xspace}
\newcommand{\HO}{H$_2$O\xspace}
\newcommand{\DO}{D$_2$O\xspace}
\newcommand{\degreesC}{$\,^{\circ}$C\xspace}
\newcommand{\pH}{$p$H\xspace}
\newcommand{\pD}{$p$D\xspace}
\newcommand{\Kd}{$K_\text{d}$\xspace}
\newcommand{\mc}[3]{\multicolumn{#1}{#2}{#3}}
\newcommand{\mco}[1]{\multicolumn{1}{c}{#1}}
\newcommand{\SNO}{{\scshape{\large{sno}}}\xspace}
\newcommand{\jour}[1]{\textnormal{#1}}
\newcommand{\diff}{\, \mathrm{d}}   
\newcommand{\aprle}{\buildrel < \over {_{\sim}}}

\begin{document}

\begin{frontmatter}

\title{\vspace{-4cm}
       Measurement of radium concentration in water with \\
       Mn-coated beads at the Sudbury Neutrino Observatory}

\author[Guelph]{T.~C.~Andersen},
\author[Carleton]{I.~Blevis},
\author[Brookhaven]{J.~Boger},
\author[Queens]{E.~Bonvin},
\author[Carleton]{M.~Chen},
\author[Oxford]{B.~T.~Cleveland}$^{,*}$,
\corauth{Corresponding author.  Address: SNO Project, PO Box 159, Lively,
Ontario  P3Y 1M3, Canada.}
\ead{bclevela@surf.sno.laurentian.ca}
\author[Oxford]{X.~Dai},
\author[Carleton]{F.~Dalnoki-Veress},
\author[Oxford]{G.~Doucas},
\author[Carleton]{J.~Farine\thanksref{Laur}},
\thanks[Laur]{Present address: Department of Physics and Astronomy,
Laurentian University, Sudbury, Ontario P3E 2C6, Canada}
\author[Oxford]{H.~Fergani},
\author[Oxford]{A.~P.~Ferraris},
\author[LosAlamos]{M.~M.~Fowler},
\author[Brookhaven]{R.~L.~Hahn},
\author[Laurentian]{E.~D.~Hallman},
\author[Carleton]{C.~K.~Hargrove},
\author[Guelph]{P.~Jagam},
\author[Oxford]{N.~A.~Jelley},
\author[Oxford]{A.~B.~Knox},
\author[Queens]{H.~W.~Lee},
\author[Carleton]{I.~Levine},
\author[Oxford]{S.~Majerus},
\author[Carleton]{K.~McFarlane},
\author[Carleton]{C.~Mifflin},
\author[LosAlamos]{G.~G.~Miller},
\author[Carleton]{A.~J.~Noble},
\author[LosAlamos]{P.~Palmer},
\author[Brookhaven]{J.~K.~Rowley},
\author[Carleton]{M.~Shatkay},
\author[Guelph]{J.~J.~Simpson},
\author[Carleton]{D.~Sinclair},
\author[Guelph]{J.-X.~Wang},
\author[LosAlamos]{J.~B.~Wilhelmy},
\and
\author[Brookhaven]{M.~Yeh}

\address[Guelph]{Physics Department, University of Guelph, Guelph, Ontario
         N1G 2W1, Canada}
\address[Carleton]{Carleton University, Ottawa, Ontario K1S 5B6, Canada}
\address[Brookhaven]{Chemistry Department, Brookhaven National Laboratory,
         Upton, New York 11973-5000, USA}
\address[Queens]{Department of Physics, Queen's University, Kingston,
         Ontario K7L 3N6, Canada}
\address[Oxford]{Nuclear and Astrophysics Laboratory, University of Oxford,
         Keble Road, Oxford, OX1 3RH, UK}
\address[LosAlamos]{Los Alamos National Laboratory, Los Alamos, New Mexico
         87545, USA}
\address[Laurentian]{Department of Physics and Astronomy, Laurentian
         University, Sudbury, Ontario P3E 2C6, Canada}

\begin{abstract}

     We describe a method to measure the concentration of \nuc{224}{Ra} and
\nuc{226}{Ra} in the heavy water target used to detect solar neutrinos at
the Sudbury Neutrino Observatory (\SNO) and in the surrounding light water
shielding.  A water volume of (50--400)~m$^3$ from the detector is passed
through columns which contain beads coated with a compound of manganese
oxide onto which the Ra dissolved in the water is adsorbed.  The columns are
removed, dried, and mounted below an electrostatic chamber into which the Rn
from the decay of trapped Ra is continuously flowed by a stream of N$_2$
gas.  The subsequent decay of Rn gives charged Po ions which are swept by
the electric field onto a solid-state $\alpha$ counter.  The content of Ra
in the water is inferred from the measured decay rates of \nuc{212}{Po},
\nuc{214}{Po}, \nuc{216}{Po}, and \nuc{218}{Po}.  The Ra extraction
efficiency is $>$95\%, the counting efficiency is 24\% for \nuc{214}{Po} and
6\% for \nuc{216}{Po}, and the method can detect a few atoms of
\nuc{224}{Ra} per m$^3$ and a few tens of thousands of atoms of
\nuc{226}{Ra} per m$^3$.  Converted to equivalent equilibrium values of the
topmost elements of the natural radioactive chains, the detection limit in a
single assay is a few times $10^{-16}$~g~Th or U/cm$^3$.  The results of
some typical assays are presented and the contributions to the systematic
error are discussed.

\end{abstract}

\end{frontmatter}

\section{Introduction}

     All experiments to detect solar neutrinos have observed significantly
fewer than are predicted by well calibrated solar models.  It is now
recognized, after publication of the first results of the \SNO
experiment~\cite{SNO123}, that this deficit is the result of the
transformation of the electron neutrino into other neutrino species.  A
crucial element of the \SNO experiment is the verification that the rates of
background processes that might be misinterpreted as neutrino events are
significantly less than the observed signal.  We describe here one of the
procedures used to measure the radioactivity content of the water in the
\SNO detector and thus to prove that the background is adequately low.

     This article is structured as follows: after giving a general
description of the \SNO experiment, the various aspects of the background
measurement technique based on Mn-coated beads are described.  Then the
methods for measuring the Ra extraction efficiency and the Po counting
efficiency are considered, followed by a discussion of systematic errors.
Finally representative results of using the method to measure the Ra
concentration in the heavy and light water of the \SNO detector are given.

\markboth {\hfill \MnOx assay method \hfill 14 January 2003}
          {14 January 2003 \hfill \MnOx assay method \hfill}

\subsection{Overview of the \SNO detector}

     The main neutrino target of the Sudbury Neutrino
Observatory~\cite{NIM00} is 1000~tonnes of \DO contained in a transparent
spherical acrylic vessel of 12~m diameter.  An array of 9438~inward-looking
photomultiplier tubes with light collectors are located on a nearly
water-tight icosahedral support structure at $\sim$3~m distance from the
central vessel.  They see the Cherenkov light from neutrino interactions in
the \DO and in part of the 1700~m$^3$ of \HO which fills the volume between
the acrylic vessel and the photomultipliers.  An additional 5700~m$^3$ of
\HO between the photomultipliers and the cavity wall provide further
shielding from external radiation.

     Solar neutrinos are detected in \SNO through three distinct reactions:
\begin{flalign*}
\nu_\text{e}\mspace{2.5mu} & + \text{d} \       \longrightarrow
        \text{e}^- \!+ \text{p} + \text{p} & \text{(CC),} \\
\nu_x                      & + \text{d} \       \longrightarrow
        \nu_x        + \text{p} + \text{n} &  \text{(NC),} \\
\nu_x                      & + \text{e}^-  \!\! \longrightarrow
        \nu_x        + \text{e}^-          &  \text{(ES),}
\end{flalign*}
where $x$ denotes any of the active neutrino species $\text{e}, \mu,
\text{or } \tau$.  The charged-current reaction (CC) observes only electron
neutrinos, the neutrino type produced in the Sun.  The neutral-current
reaction (NC) detects the total flux of active neutrinos and has equal
sensitivity to all flavors.  The elastic-scattering reaction (ES) also
detects all active neutrinos, but is dominantly sensitive to neutrinos of
the electron type.

\subsection{Upper limits on radioactive contamination}

     The most restrictive limits on \DO purity arise by considering the
background for the neutral current reaction.  The main unvetoed background
for this reaction is the photodisintegration of the deuteron,
$\text{d}(\gamma,\text{p})\text{n}$, whose threshold is 2.22~MeV.  The only
commonly occurring natural isotopes which emit $\gamma$ rays above this
energy are near the end of the Th and U decay chains, viz., \nuc{208}{Tl},
99.8\% of whose decays produce a 2.614-MeV $\gamma$ ray, and \nuc{214}{Bi},
2.15\% of whose decays are accompanied by a $\gamma$ ray with energy greater
than 2.22~MeV.  If we require that the rate of neutron production by
radioactive background be less than 1 per day, which is less than 10\% of
the anticipated signal, then, based on our Monte Carlo for the response of
the detector to internal $\gamma$ rays, we set the goal of restricting the
decay rate of \nuc{208}{Tl} and \nuc{214}{Bi} in the \DO of the \SNO
detector to less than 480/day and 32~000/day, respectively.  Assuming
radioactive equilibrium in the Th and U chains, this leads to upper limits
of 4.2\E{-15} g~Th/cm$^3$~\DO and 3.3\E{-14} g~U/cm$^3$~\DO.  \footnote{It
is of course not true that the Th and U chains are in equilibrium in the
water of the \SNO detector all the way back to long-lived \nuc{232}{Th} and
\nuc{238}{U}, but it is conventional in low-background counting to express
radioactivity measurements as if equilibrium were present, and we follow
this practice here.}  At these limits there is an average of 1.5~atoms of
\nuc{208}{Tl} and 640~atoms of \nuc{214}{Bi} in the entire \DO volume.

     The requirements on radiochemical purity of the \HO in the \SNO
detector are determined by the ingress of $\gamma$ rays into the \DO region
that are produced by the decay of \nuc{208}{Tl} or \nuc{214}{Bi} in the
surrounding light water.  Based on our early simulations for the transport
of $\gamma$ rays and assuming equilibrium, this leads to upper limits of
3.7\E{-14} g~Th/cm$^3$~\HO and 4.5\E{-13} g~U/cm$^3$~\HO for the water in
the region between the acrylic vessel and the photomultiplier tubes,
considerably less severe than required for the \DO.

     Four complementary techniques have been developed by \SNO to determine
if the detector water meets these specifications.  Three methods are based
on flowing water to an external extraction system where a chemical
separation of Ra, Th, or Rn is performed; the other method uses the
Cherenkov light signals from the photomultipliers to directly infer the
concentrations of \nuc{214}{Bi} and \nuc{208}{Tl}.  The first of the
extraction methods, the \MnOx assay method, is the subject of this paper.
The second radiochemical method, which is described in a companion paper
\cite{HTiO}, flows water over a filter coated with an adsorber of hydrous
titanium oxide.  After water flow, the filter is eluted with acid to remove
the extracted Ra, Th, and Pb.  These elements are then concentrated, mixed
with liquid scintillator, and the $\beta$--$\alpha$ coincidences of
\nuc{212}{Bi}--\nuc{212}{Po} and \nuc{214}{Bi}--\nuc{214}{Po} are detected
with a photomultiplier.  In the third chemical assay method, water from the
detector is flowed through a degasser to liberate Rn.  The Rn is purified
and collected and its alpha decays are counted in a Lucas cell scintillator
chamber on a photomultiplier~\cite{Liu92}.  This method can detect only the
U-chain isotope \nuc{222}{Rn}.

\section{The \MnOx assay method}

     In overview, water is passed through columns that contain beads coated
with a manganese oxide compound.  The coating extracts Ra from the flowing
water, and, to a lesser extent, other dissolved species, such as Th, Pb,
etc. \cite{Moore}.  After a large volume of water has passed through the
columns, they are removed and dried.  The Rn produced from Ra decay is swept
from the columns into an electrostatic chamber where it decays.  The charged
Po ions from the decay of Rn are carried by an electric field onto an alpha
counter where the decays of the Po are detected.

     In this section we describe all the components of the \MnOx assay
method as used in \SNO, from the \MnOx coating onto acrylic beads, the
column which holds the beads during an assay, the system that flows water
through the column, the apparatus for counting the column, the physical
interpretation of the data, and the way in which the data are analyzed.

\subsection{Beads and Mn coating}
\label{beads}

     Solid spherical acrylic beads of 600~$\mu$m nominal diameter were
chosen as the support material.  These beads have good mechanical strength,
little water uptake, low intrinsic content of radioactive
elements~\cite{Guelph}, good resistance to the production of fine
particulate material (fines), and good Rn emanation efficiency.

     Tests of Ra extraction efficiency and retention, Rn emanation, and
fines production were carried out with two forms of coatings, designated
MnO$_2$ and \MnOx.  The MnO$_2$ coating was produced by prolonged oxidation
of the beads with sodium permanganate in the presence of sulfuric acid at
high temperature.  This yields a black coating on the bead surface.  The
\MnOx coating is obtained by incomplete oxidation with the same reagents,
and gives a dark brown coating.  It was selected for our purpose as the Rn
emanation efficiency was higher than with MnO$_2$.  The exact stoichiometry
has not been determined so we call it \MnOx.

     The procedure for bead production with \MnOx coating is described in
\cite{vers1}.  Electron microscopy reveals that the coating consists of a
base layer that is parallel to the surface and needles of (0.3--0.5)~$\mu$m
length oriented approximately perpendicular to the surface.  The average
coating density is $0.45\pm0.1$~g~Mn/m$^2$.  The radioactive background of
one sample of coated beads was measured with a Ge $\gamma$-ray
detector~\cite{Jagam} to be $<$90~ng~Th/g~beads and $29\pm14$~ng~U/g~beads.

     The ability of the beads to retain adsorbed Ra was measured by spiking
a column with \nuc{226}{Ra}, measuring its initial activity $A(0)$, flowing
a volume of water $V$ through the column, and measuring the activity $A(V)$
that remained on the column.  These tests were conducted on small-scale
analogues of our columns used in extraction with flow rates of
20~column-volumes per min and the activity was measured with a Ge
$\gamma$-ray detector.  The results of these measurements can be described
in terms of an effective distribution coefficient \Kd defined as
\begin{equation}
\label{Kd}
K_\text{d}(V) = \frac{V/V_\text{beads}}
                     {A(0)/A(V) - 1},
\end{equation}
where $V_\text{beads}$ is the volume of the beads in the column.  Up to
$V/V_\text{beads} = 170~000$, \Kd was found to be $>$10$^6$ with an error of
approximately 25\%.  This implies there is little concern regarding loss of
Ra from a one liter column at a flow rate of 20~$\ell$/min\footnote{To avoid
confusion with the number 1, we use the symbol ``$\ell$'' as an abbreviation
for `liter.'} up to water volumes of at least 170~k$\ell$.

     Not only do the beads adsorb Ra, but they also have an affinity for
other elements, such as Ba.  By flowing a large volume of Ba solution over a
fixed volume of \MnOx-coated beads and measuring the Ba concentration of the
eluate, the capacity of the coated beads to hold Ba was determined to be
$\sim$300~mg~Ba/$\ell$~beads.  The capacity for Ra should be comparable as
it also is a divalent atom and has only slightly greater radius.  The other
divalent atoms and monovalent atoms have a smaller affinity for \MnOx.  The
water in the \SNO detector does not contain such large quantities of
impurities that there is any problem with the bead capacity.

\subsection{Column}

     During a water assay and subsequent counting, the coated beads are held
within a column made from polypropylene.  The internal column volume is
$\sim$950~cm$^3$ and it is filled with $700.0\pm0.2$~g of coated beads which
contain approximately 3.5~g of Mn.

     Before a column is used it is dried and counted for background.  This
is done in the same way as after the column has been used in an experiment
and will be described below.  We call this measurement the column `blank.'
When counting is finished, which takes at least 10~days, the column is
filled with nitrogen, its ends are capped, and it is brought underground for
the assay.

     To maintain the isotopic purity of the heavy water, columns for a \DO
assay are deuterated before use.  This is done by flowing $\sim$3~$\ell$ of
\DO through the column with a peristaltic pump at 150~cm$^3$/min until the
density of the water that exits the column is $>$1.103~g/cm$^3$.  (The
density of pure \DO at 20\degreesC is 1.105~g/cm$^3$.)  To reduce the entry
of fines into the water, columns are then rinsed at a rate of 20~$\ell$/min
for 15 min.

\subsection{Water flow}

     Most \HO assays use a sampling point halfway between the acrylic vessel
and the photomultipliers as it is this region of \HO that contributes most
to the neutral-current background.  Assays of the \DO usually draw water
from the vessel bottom and return it near the vessel top.  The \pH of the
\HO is 5.8--6.0 and the \pD of the \DO is 5.5--5.8.  All water entering the
detector is cooled; the temperature of the \DO and the immediately
surrounding \HO is (10.5--11)\degreesC.

     After the column is attached to the water system it is evacuated so
that the ingress of \nuc{222}{Rn} to the system is minimized.  The usual
water flow rate through an \MnOx column is 20~$\ell$/min.  To allow the
sampling of very large volumes of water, and thus to achieve greater
sensitivity, assays of the \DO are usually made with 4~columns through which
the water flows in parallel with 20~$\ell$/min through each.

     All water that has flowed through an \MnOx column is passed through
ultrafilters that remove tiny particulates (fines).  If \MnOx fines were to
be carried into the acrylic vessel, they would bind there with Ra and remove
it from future assays, thus leading to an underestimation of the true Ra
content of the water.  The ultrafilration units have a 3~kiloDalton
molecular weight cutoff (approximately 3~nm).  Their permeate flow returns
to the water system; their concentrate flow is passed through a 0.1~$\mu$m
filter and then returned to the feed stream of the \MnOx column.

     Since a very large volume of water is passed through the column, there
is a concern that some Mn may be dissolved and enter the water.  An
indication that some loss of Mn occurs is provided by the visual observation
that the bead color sometimes changes from its initial dark brown to light
brown after water flow.  This effect is most evident on the end of the
column where the water enters.  To address this question, experiments have
been made in which \HO at \pH~7 was recirculated from a 1~m$^3$ tank through
a column and the concentration of Mn measured as a function of flow volume
by inductively-coupled plasma mass spectrometry \cite{Seastar}.  Very little
Mn was found, even up to water volumes of 115~m$^3$.  Using this data and
measurements of the mass of Mn remaining on the beads after their use in an
extraction, it appears that a few tens to hundreds of mg of Mn are removed
from a column in a large volume assay.  Most of this Mn is in the form of
fines which are blocked from entering the detector by the ultrafilter that
follows the \MnOx columns.  Very little of the released Mn is in dissolved
form: measurements of the \DO in the acrylic vessel show an increase in Mn
concentration from an initial value of 0.5~ng/cm$^3$ to 2.0~ng/cm$^3$ during
the pure \DO phase, a period in which 61~\MnOx columns were used in assays.
This increase in Mn concentration has not deleteriously affected the
operation of the \SNO detector.

\subsection{Column drying and counting}

     When an assay has been completed, the columns are removed, their ends
are capped, and they are brought to a surface laboratory where they are
dried and counted.  Either air or boil-off gas from liquid N$_2$ is used for
drying.  The gas is filtered and heated to 60\degreesC.  A volume of
$\sim$25~m$^3$ is used to dry a wet column from an assay; $\sim$10~m$^3$ is
sufficient for an initially dry column, as when counting a blank before use
in an assay.

\begin{figure}[!t]
\includegraphics[width=\hsize]{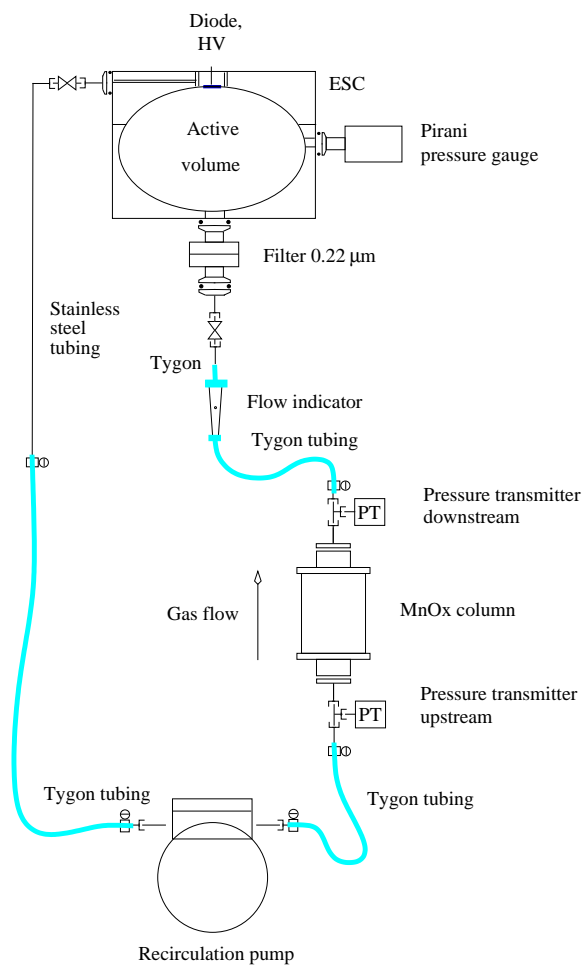}
\caption{Schematic diagram of gas flow through \MnOx column and ESC during
counting.}
\label{escloop}
\end{figure}

     The Ra activity on the columns is measured by attaching the dried
column to the gas flow loop on an electrostatic counter (ESC), as shown in
Fig.~\ref{escloop}.  To remove Rn and residual water vapor, the counting
loop is initially filled with dry N$_2$ from a compressed gas bottle and
evacuated several times.  The loop is then pressurized to $26\pm1$~mbar with
dry N$_2$ which acts as a carrier gas to transport Rn from the column to the
ESC.  The total elapsed time between the end of extraction and the start of
counting is typically (10--12)~hours.

     The internal shape of the ESC chamber was designed~\cite{NIM99} to
sweep the positively charged Po ions from Rn decay onto an alpha detector at
the apex of the chamber with high efficiency throughout its volume.  The
alpha counter is an 18~mm by 18~mm windowless silicon photodiode operated at
a bias voltage of (60--69)~volts.  The diode is held at a potential of
1000~volts below that of the aluminum chamber.  The output of the diode is
amplified, digitized in 1024~channels, and written to disk every 3~hours.
Counting is continued for (10--30)~days.  Long counting times are essential
in \DO assays to well determine the \nuc{228}{Th} activity.

     Fig.~\ref{escloop} shows the counting configuration for a single-column
assay.  When 4~columns are used they are attached to the gas flow loop in
parallel in the same position as with one column.  The gas is circulated
with a diaphragm pump at a typical flow rate of (0.3--0.4)~STP $\ell$/min.
The filter just below the ESC blocks the entry of fines to the chamber and
is the major impediment to gas flow.

     Most of the Ra from a water assay is located near the top of the column
where the water enters.  To give the highest efficiency for the transport of
Rn to the active ESC volume, especially the short-lived \nuc{220}{Rn}, the
gas flow through the column is thus opposite to the previous direction of
water flow.

\begin{figure}[!t]
\includegraphics*[bb=13 72 476 367,width=\hsize]{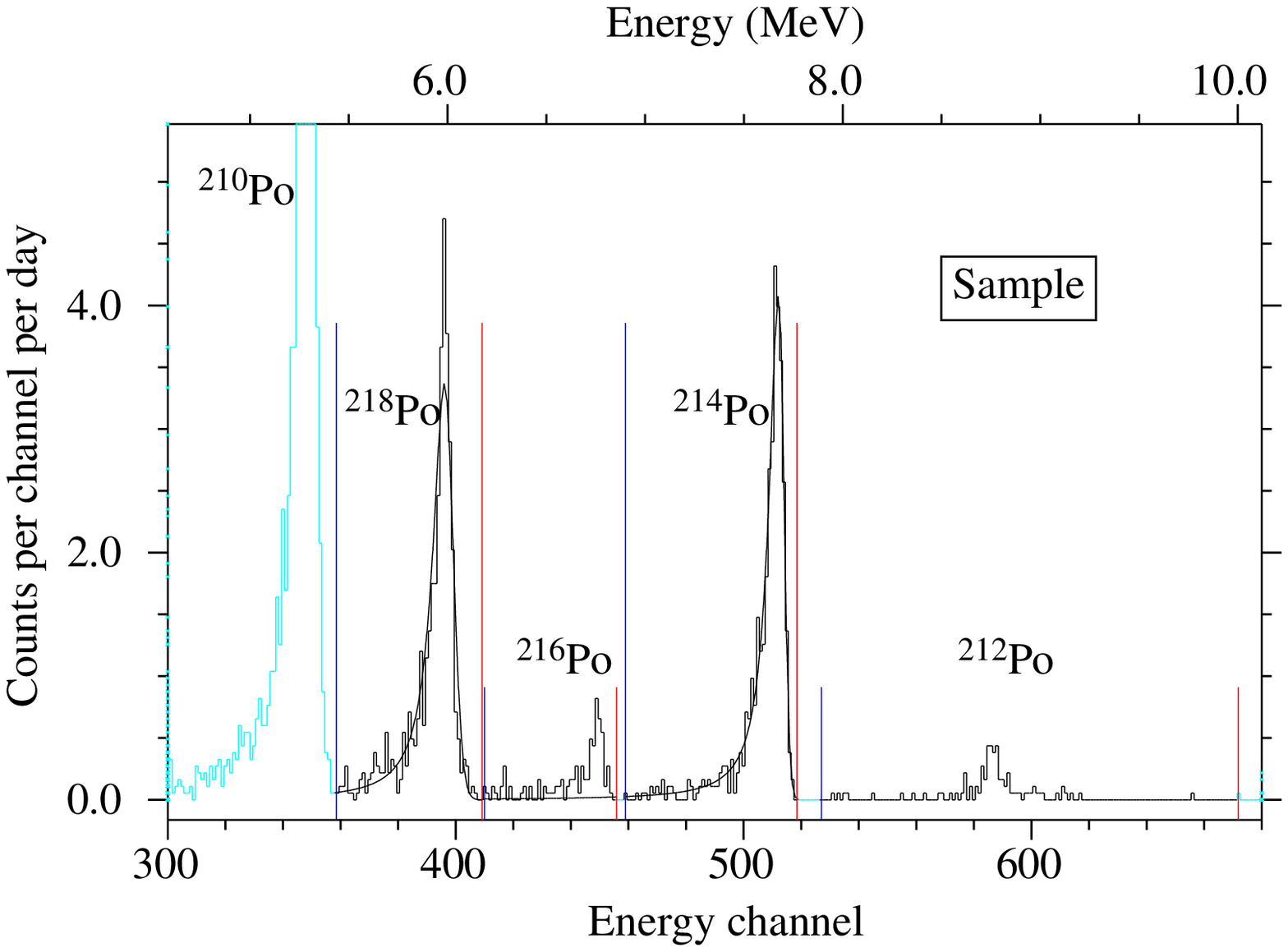}
\includegraphics*[bb=13 26 476 324,width=\hsize]{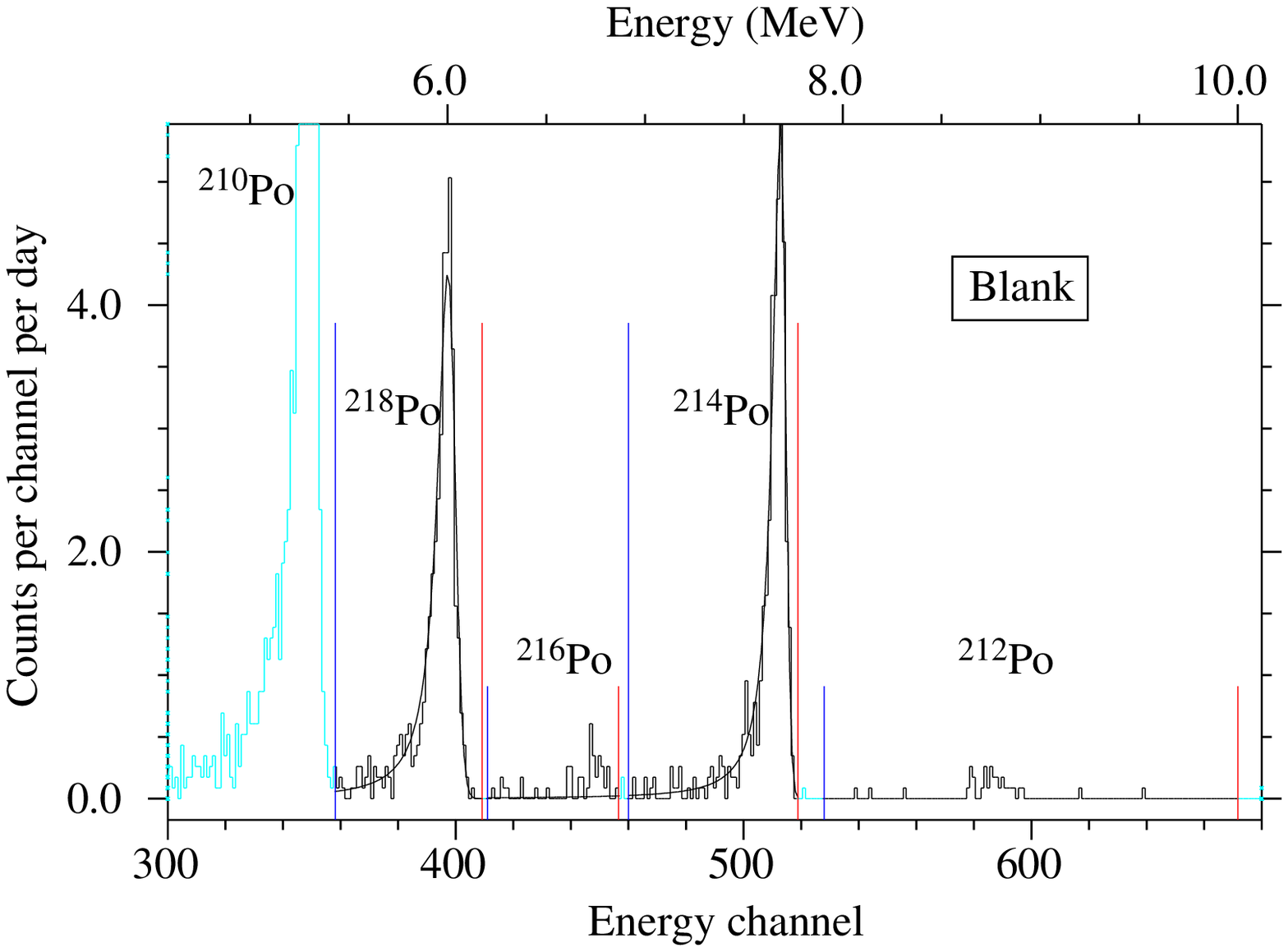}
\caption{Typical energy spectra from $\alpha$ detector at the end of
counting of a 4-column \DO assay.  Lower panel: before assay, the `blank.'
Upper panel: after passing 355~k$\ell$ of \DO through columns, the `sample.'
The peaks in order of increasing energy are \nuc{210}{Po} at 5.3~MeV,
\nuc{218}{Po} at 6.0~MeV, \nuc{216}{Po} at 6.8~MeV, \nuc{214}{Po} at
7.7~MeV, and \nuc{212}{Po} at 8.8~MeV.  The vertical lines show the energy
windows assigned for each peak.}
\label{readexample}
\end{figure}

     Energy spectra from the $\alpha$ detector at the end of counting of a
typical assay are given in Fig.~\ref{readexample}.  The lower panel shows
the spectrum of the blank before passing water through the column; the upper
panel shows the spectrum of the sample after water flow.  Because the
$\alpha$ particles lose energy as they pass through the surface dead layer
of the diode, each Po peak falls gradually on the low-energy side, but much
more steeply on the high-energy side.

\subsection{Interpretation of energy spectra}

     Since it is Po produced by the decay of Rn that is detected, we look
backward from Rn along the Th and U chains to the last isotope whose half
life is considerably longer than the usual (10--15)~day counting period.
This leads to 1.9-y \nuc{228}{Th} in the Th chain and 1600-y \nuc{226}{Ra}
in the U chain.  If any prior elements in either chain are present, their
influence is effectively blocked by these long-lived isotopes.

     The relevant part of the Th decay chain and the energies of the
$\alpha$ particles that follow Rn is thus
  \nuc{228}{Th} $\xrightarrow[\text{1.91 y}]{\alpha}$
  \nuc{224}{Ra} $\xrightarrow[\text{3.66 d}]{\alpha}$
  \nuc{220}{Rn} $\xrightarrow[\text{55.6 s}]{\alpha}$
  \nuc{216}{Po} $\xrightarrow[\text{0.15 s}]{\text{6.8 MeV }\alpha}$
  \nuc{212}{Pb} $\xrightarrow[\text{10.6 h}]{\beta}$
  \nuc{212}{Bi} $\xrightarrow[\text{1.01 h}]{\beta}$
  \nuc{212}{Po} $\xrightarrow[\text{298 ns}]{\text{8.8 MeV }\alpha}$
  \nuc{208}{Pb}, which is stable.  The next to last decay here takes place
64\% of the time; the other 36\% proceed through
  \nuc{212}{Bi} $\xrightarrow[\text{1.01 h}]{\text{6.1 MeV }\alpha}$
  \nuc{208}{Tl} $\xrightarrow[\text {3.0 m}]{\beta}$
  \nuc{208}{Pb}.  The major peaks from Th are thus \nuc{216}{Po} and
\nuc{212}{Po}, as labeled in Fig.~\ref{readexample}.  The 6.1-MeV $\alpha$
from \nuc{212}{Bi} makes an unresolved peak near the upper edge of the much
stronger \nuc{218}{Po} peak.  The window for \nuc{212}{Po} decay is set very
wide to include the detection of the 2.2-MeV endpoint energy electron from
\nuc{212}{Bi} decay in coincidence with the 8.8-MeV $\alpha$ from
\nuc{212}{Po} decay.

     The section of the U chain of interest is
  \nuc{226}{Ra} $\xrightarrow[\text{1600 y}]{\alpha}$
  \nuc{222}{Rn} $\xrightarrow[\text{3.82 d}]{\alpha}$
  \nuc{218}{Po} $\xrightarrow[\text{3.10 m}]{\text{6.0 MeV }\alpha}$
  \nuc{214}{Pb} $\xrightarrow[\text{26.8 m}]{\beta}$
  \nuc{214}{Bi} $\xrightarrow[\text{19.9 m}]{\beta}$
  \nuc{214}{Po} $\xrightarrow[\text{162 $\mu$s}]{\text{7.7 MeV }\alpha}$
  \nuc{210}{Pb}.  The chain is completed by
  \nuc{210}{Pb} $\xrightarrow[\text{22.3 y}]{\beta}$
  \nuc{210}{Bi} $\xrightarrow[\text {5.0 d}]{\beta}$
  \nuc{210}{Po} $\xrightarrow[\text {138 d}]{\text{5.3 MeV }\alpha}$
  \nuc{206}{Pb}, which is stable.  Fig.~\ref{readexample} shows the major
peak regions due to \nuc{218}{Po} and \nuc{214}{Po} in a U-dominated
spectrum.  Although the \nuc{210}{Po} peak is also a part of the U chain, it
cannot be used in analysis as it represents the gradual accumulation on the
diode of long-lived \nuc{210}{Pb} from the counting of previous assays,
tests, etc.

     Because the concentration of \nuc{226}{Ra} as an impurity in the ESC
loop is higher than that of \nuc{228}{Th}, and the counting efficiency for
\nuc{222}{Rn} is higher than that of \nuc{220}{Rn}, all low-level spectra
are U-dominated, rather than Th-dominated.  Thus, the relative heights of
the peaks in a single spectrum do not reflect the relative concentrations of
\nuc{226}{Ra} and \nuc{224}{Ra} in the water.

     Note in Fig.~\ref{readexample} that the \nuc{216}{Po} and \nuc{212}{Po}
peaks are somewhat higher in the sample spectrum than in the blank spectrum.
This increase occurs because the column extracted \nuc{224}{Ra} from the
flowing water.  In contrast, the \nuc{218}{Po} and \nuc{214}{Po} peaks are
higher in the blank than in the sample.  This difference is not because
significant \nuc{226}{Ra} was lost during water flow; rather, it is due to
the presence of more \nuc{222}{Rn} in the ESC loop when counting of the
blank was initiated than when counting of the sample began.

\subsection{Data analysis}

     The counting data are analyzed in three steps: In the first step, the
energy spectra are fit, energy windows are set about the four Po peaks, and
the number of counts in each energy window is determined as a function of
time.  The second step is to fit these time spectra to determine the decay
rates of the isotopes that produce the counts.  These two steps are done
separately for the blank and for the sample.  In the third step, the isotope
decay rates for sample and blank and the water flow rate are used to
determine the concentration of \nuc{224}{Ra} and \nuc{226}{Ra} in the water.

\begin{table}[t]
\caption{Data, fitted, and calculated parameters for spectra in
Fig.~\protect\ref{readexample}.}
\label{readparameters}
\begin{tabular*}{\hsize}{@{} l @{\extracolsep{\fill}} @{\hspace{-0.7em}} d
@{\hspace{-0.5em}} d @{}}
\hline
\hline
                &   \mc{2}{c}{Value}  \\ \cline{2-3}
Item            &   \mco{Blank}   &   \mco{Sample}  \\
\hline
Live time (days)                &  11.522  &  18.288  \\
Counts in \nuc{218}{Po} window  & 485      & 683      \\
Counts in \nuc{216}{Po} window  &  45      & 122      \\
Counts in \nuc{214}{Po} window  & 475      & 641      \\
Counts in \nuc{212}{Po} window  &  32      &  93      \\
Peak height $H$ (counts)      \\ \hspace{1em}
\nuc{218}{Po}       &  49.3^{+3.5}_{-3.3}   &  61.6^{+3.7}_{-3.5}  \\
                             \hspace{1em}
\nuc{214}{Po}       &  63.0^{+4.5}_{-4.3}   &  74.9^{+4.7}_{-4.4}  \\
Peak location $C$ (channels)  \\ \hspace{1em}
\nuc{218}{Po}       & 397.4^{+0.3}_{-0.3}   & 396.1^{+0.3}_{-0.3}  \\
                             \hspace{1em}
\nuc{214}{Po}       & 513.0^{+0.3}_{-0.3}   & 512.3^{+0.3}_{-0.3}  \\
Width of Lorentzian $W_\text{L}$ (keV) \\ \hspace{1em}
below \nuc{218}{Po} & 134.7^{+11.2}_{-10.3} & 145.1^{+10.3}_{-9.5} \\
                             \hspace{1em}
below \nuc{214}{Po} & 105.8^{+ 8.4}_{- 7.8} & 123.5^{+ 8.4}_{-8.0} \\
Width of Gaussian $W_\text{G}$ (keV)   \\ \hspace{1em}
above \nuc{218}{Po} &  84.7^{+ 7.4}_{- 6.8} & 102.2^{+ 6.5}_{-6.1} \\
                             \hspace{1em}
above \nuc{214}{Po} &  57.0^{+ 6.1}_{- 5.7} &  60.2^{+ 6.0}_{-5.6} \\
Overlap of \nuc{214}{Po} into \nuc{216}{Po} (\%) &   1.4  &   1.7  \\
Window efficiency for \nuc{218}{Po} (\%)         &  95.5  &  95.0  \\
Window efficiency for \nuc{216}{Po} (\%)         &  96.2  &  95.4  \\
Window efficiency for \nuc{214}{Po} (\%)         &  97.2  &  96.6  \\
Window efficiency for \nuc{212}{Po} (\%)         &  98.0  &  98.0  \\
\hline
\hline
\end{tabular*}
\end{table}

\subsubsection{Energy spectrum analysis}

     The final accumulated spectrum is first passed through a peak
recognition algorithm which identifies the various peaks.  Once the peaks
are recognized, a fit is made to the two principal peaks, either
\nuc{212}{Po} and \nuc{216}{Po} for a Th-dominated spectrum or \nuc{214}{Po}
and \nuc{218}{Po} for a U-dominated spectrum.  The energy scale is then set
by a linear interpolation from the two principal peak positions and the four
energy windows are set based on the fitted peak parameters.  Finally, the
number of counts in each window in each 3-hour interval is written to a file
for subsequent analysis.

     This process is illustrated with the sample energy spectrum in
Fig.~\ref{readexample}.  The principal peaks are \nuc{214}{Po} and
\nuc{218}{Po}, so this is a U-dominated spectrum.  The solid line is the fit
to the \nuc{214}{Po} and \nuc{218}{Po} peaks.  Table~\ref{readparameters}
gives the fitted parameters and other data for these spectra.  The
\nuc{218}{Po} window is set so that its lower limit removes the
\nuc{210}{Po} peak and its upper limit fully includes the \nuc{212}{Bi}
contribution.

     This figure also shows one practical complication: although the
$\alpha$ resolution is good, it is not high enough to totally separate the
peaks.  As a result, each peak makes a contribution to the peaks at lower
energy~\cite{NIM99}.  The extension of the fitted line in
Fig.~\ref{readexample} below the \nuc{214}{Po} peak indicates the overlap of
\nuc{214}{Po} into the \nuc{216}{Po} energy window.  Usually (1--2)\% of the
counts in each window appear as overlap counts in the adjacent lower energy
window.  Although this fraction is small, overlap can be an appreciable
effect.  For example, since the \nuc{214}{Po} peak of the spectrum of the
blank in Fig.~\ref{readexample} contains 475 counts and the overlap is
1.1\%, approximately 5 of the counts in the \nuc{216}{Po} window are from
overlap, more than 10\% of the total of 45 counts in this window.

     A fit function that was found to adequately approximate the shape of
each Po peak is a Gaussian above the maximum and a Lorentzian below the
maximum.  If the peak height is $H$ and the center (maximum) is at $C$, the
number of counts $m$ at energy $E$ is then
\begin{equation}
  m(E) = H \left\{
              \begin{array}{l l}
                 \exp\biggl[-\frac{
                                    {\displaystyle 1}
                                  }
                                  {
                                    {\displaystyle 2}
                                  }
                            \biggl(
                            \frac{
                                   {\displaystyle E - C}
                                 }
                                 {
                                   {\displaystyle W_\text{G}}
                                 }
                           \biggr)^2
                     \biggr]
                                                   & \text{for }E \ge C \\
                 \frac{
                        {\displaystyle (W_\text{L}/2)^2}
                      }
                      {
                        {\displaystyle (E - C)^2 + (W_\text{L}/2)^2}
                      }
                                                   & \text{for }E \le C,
              \end{array}
           \right.
\end{equation}
where $W_\text{G}$ is the width (standard deviation) of the Gaussian and
$W_\text{L}$ is the width (FWHM) of the Lorentzian.  Based on the Poisson
probability $(m|l)$ for detecting $m$ counts in an energy bin that contains
$l$ counts, $(m|l) = e^{-m} m^l/l!$, the likelihood function $\mathcal{L}$
is defined as
\begin{equation}
\label{LF}
  \mathcal{L} = \prod_i e^{-m(E_i)} \frac{m(E_i)^{l(E_i)}}{l(E_i)!},
\end{equation}
where the product is over all the energy bins $i$ in the window.  In
practice the fit is made by searching for the set of parameters that
minimize the negative logarithm of $\mathcal{L}$,
\begin{equation}
\label{lnLF}
 -\ln \mathcal{L} = \sum_i[m(E_i) - l(E_i)\ln m(E_i)],
\end{equation}
where the constant term $\ln[l(E_i)!]$ has been neglected.  The two peaks
are fit independently and three passes are made through the fit routine,
with each pass gradually refining the four peak parameters $H, C,
W_\text{G},$ and $W_\text{L}$, and resetting the window limits.  After the
last pass, 68\% confidence regions on the parameters are calculated by
finding the values of each parameter that increase $-\ln \mathcal{L}$ by
0.5, all other parameters being maximized.

\begin{figure*}[!t]
\includegraphics*[bb=22 72 476 368,width=0.48\hsize]{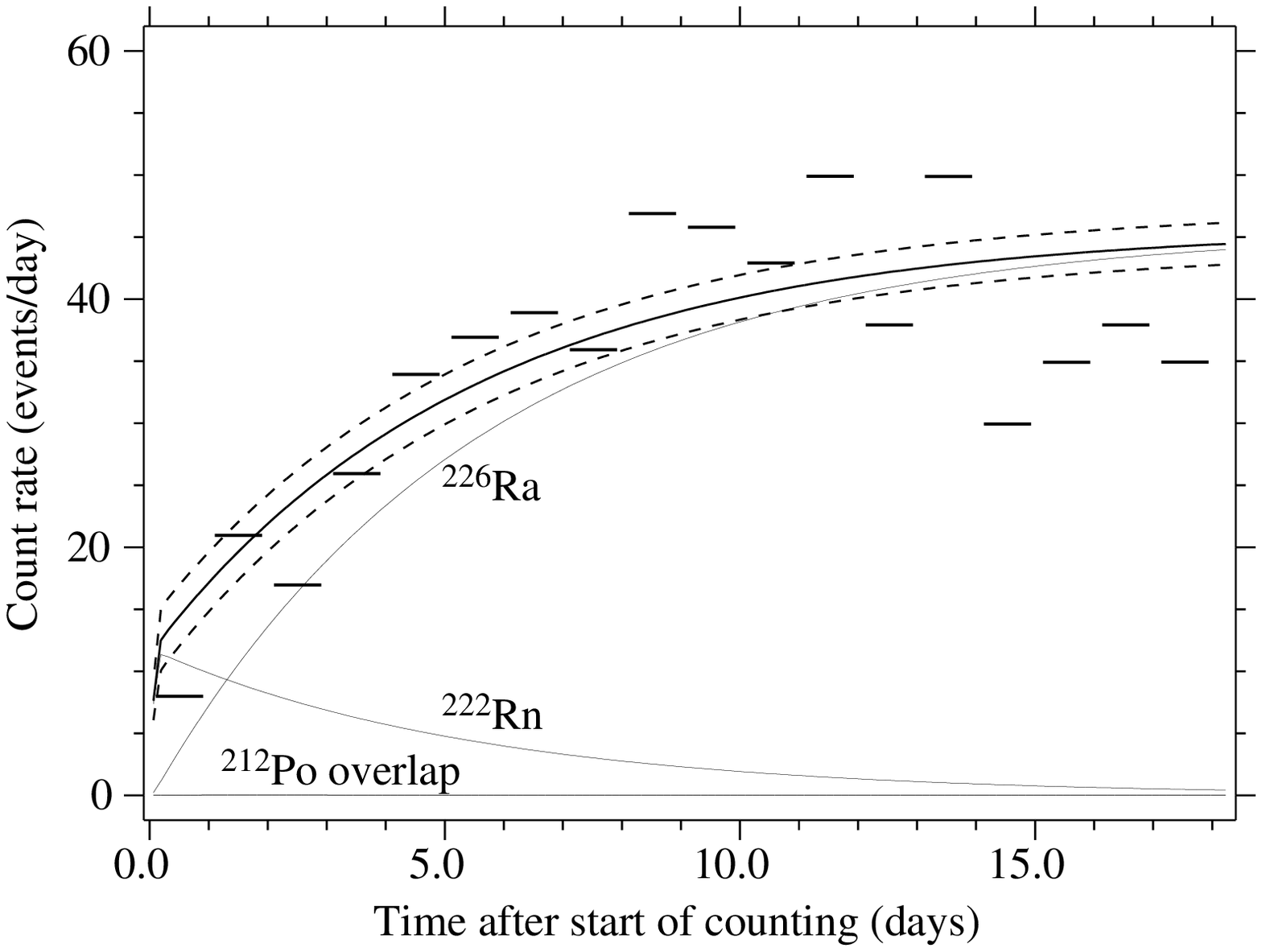}
\hfill
\includegraphics*[bb=22 72 476 368,width=0.48\hsize]{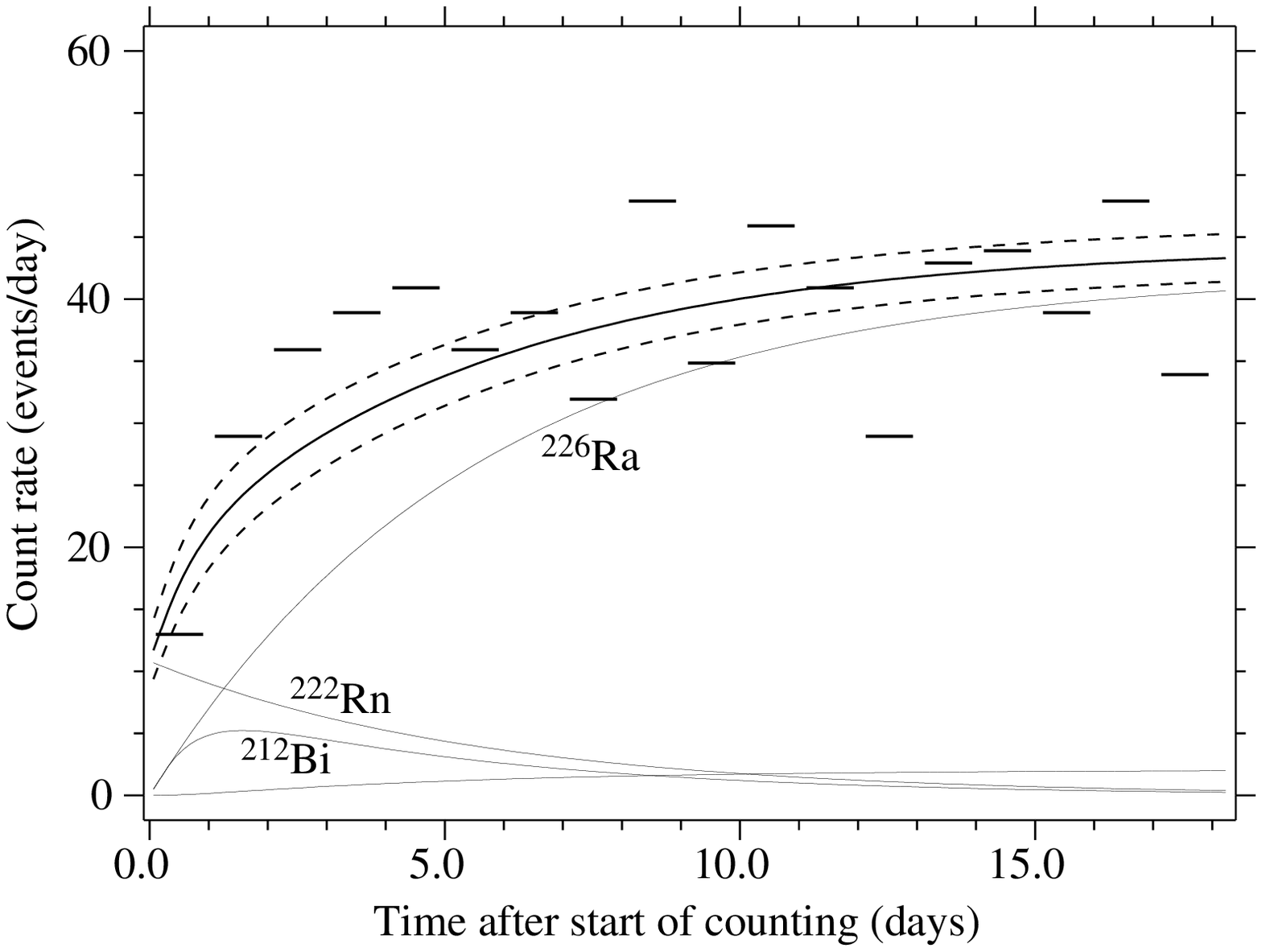}
\includegraphics*[bb=22 26 476 360,width=0.48\hsize]{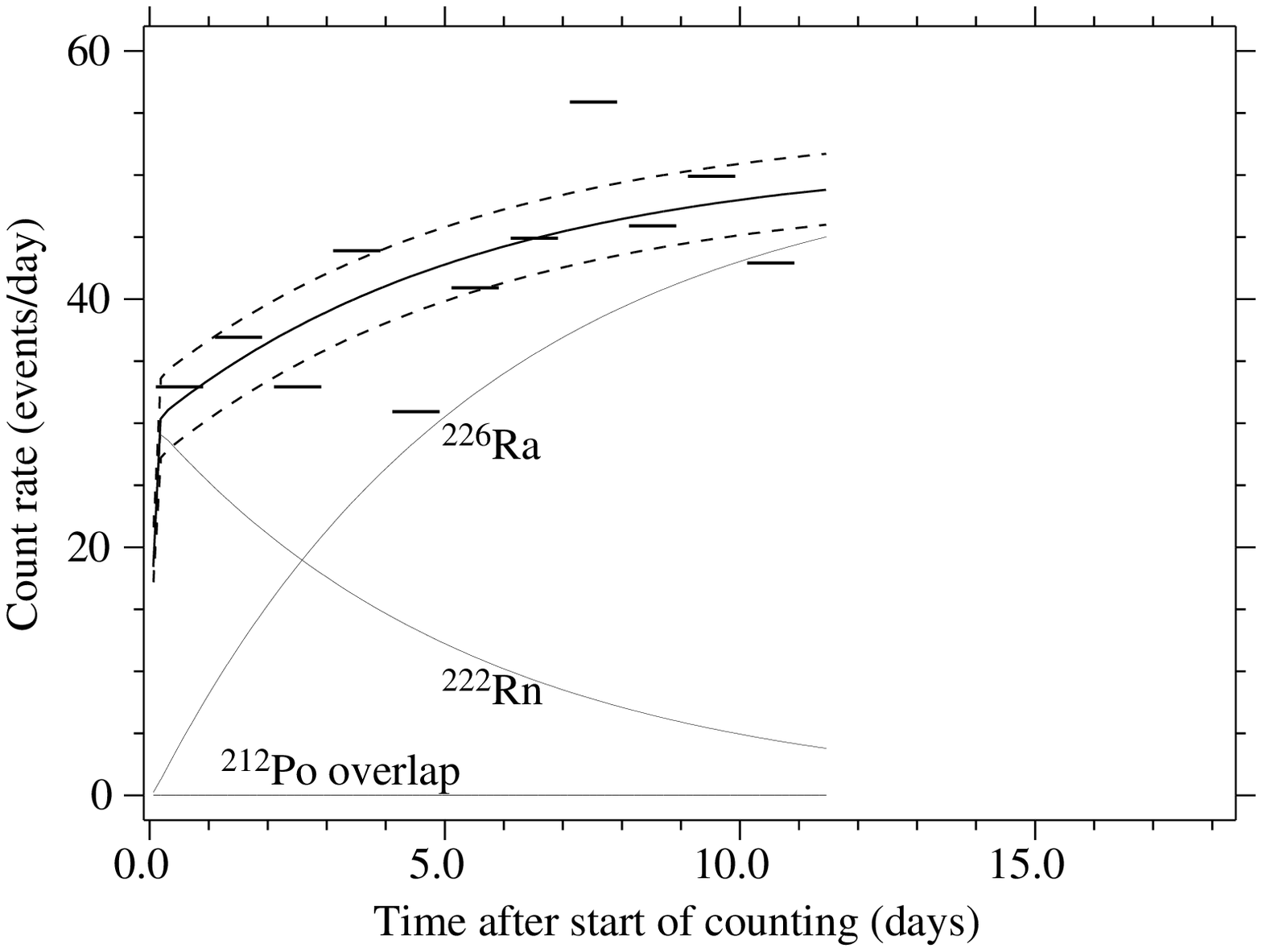}
\hfill
\includegraphics*[bb=22 26 476 360,width=0.48\hsize]{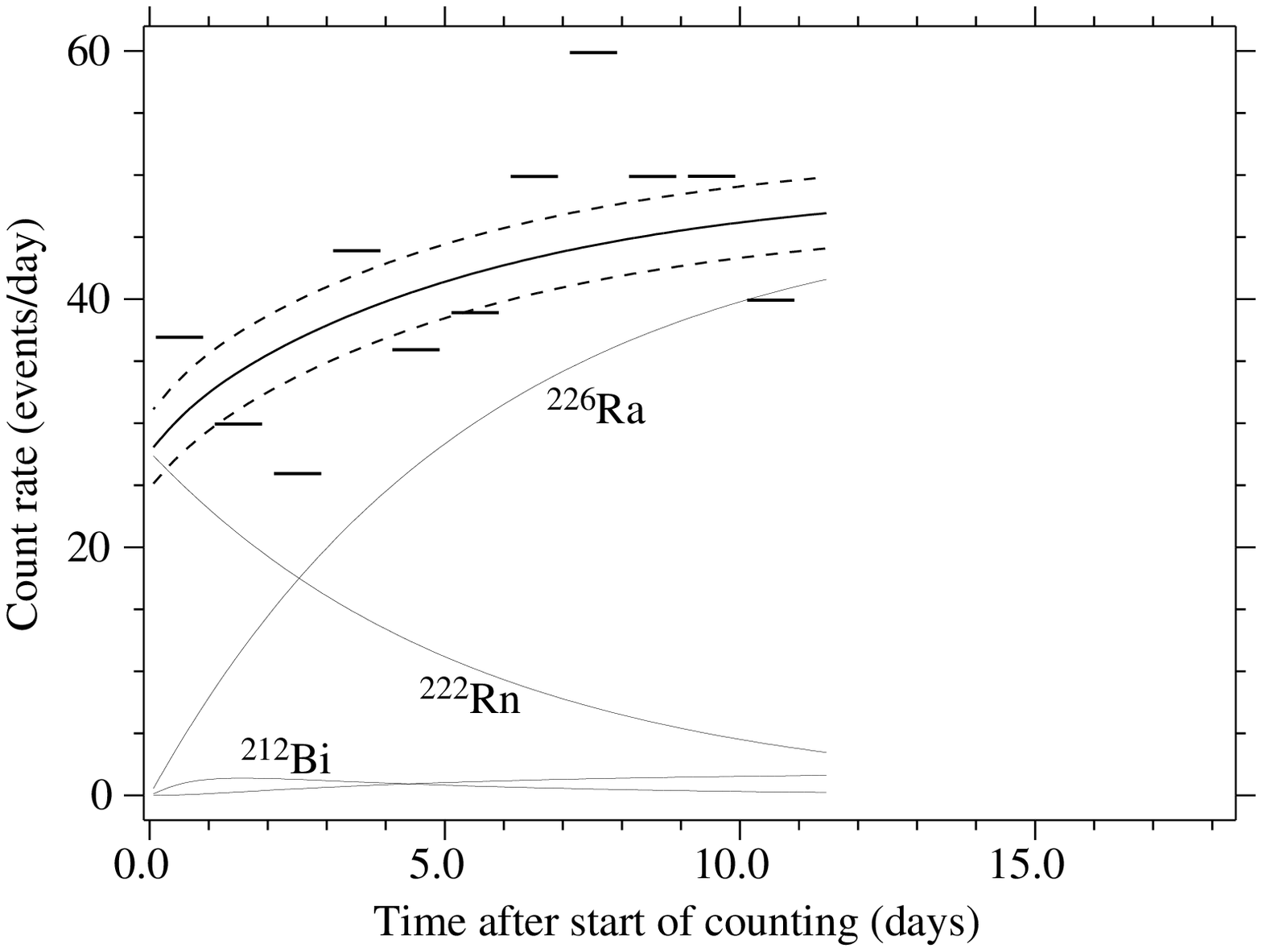}
\caption{Time spectra of \nuc{214}{Po} (left side) and \nuc{218}{Po} (right
side) for the same 4-column \DO assay as in Fig.~\protect\ref{readexample}.
Lower panels: blank before assay.  Upper panels: sample after passing \DO
through columns.  The data points are indicated by horizontal lines of 1-day
duration.  Since the count rate is very low, the data from every eight
3-hour data collection intervals have been combined.  The thick solid line
is the fit to the raw data with the encompassing dashed band indicating the
approximate 68\% confidence range.  The major components of the fit are
shown by the thin solid lines.  See text for further explanation.}
\label{fitUexample}
\end{figure*}

     Finally, the number of counts in each of the 4 energy windows for each
time interval is determined, the best fit parameters are used to calculate
the overlap and energy efficiency for each window, and all of these
calculated values are written to an output file.

\subsubsection{Time spectrum analysis}
\label{timespectra}

\begin{figure*}[!t]
\includegraphics*[bb=5 72 476 368,width=0.48\hsize]{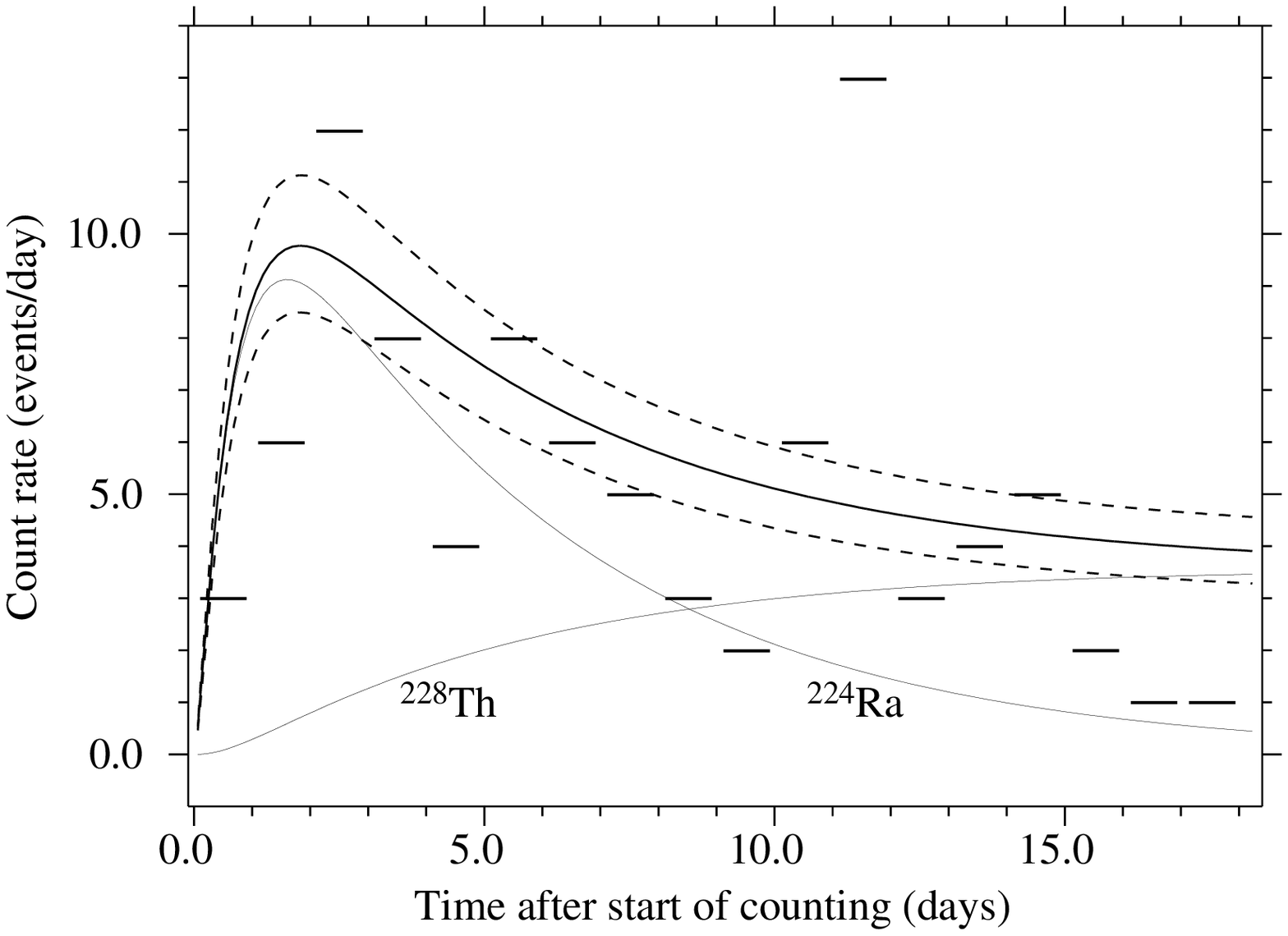}
\hfill%
\includegraphics*[bb=5 72 476 368,width=0.48\hsize]{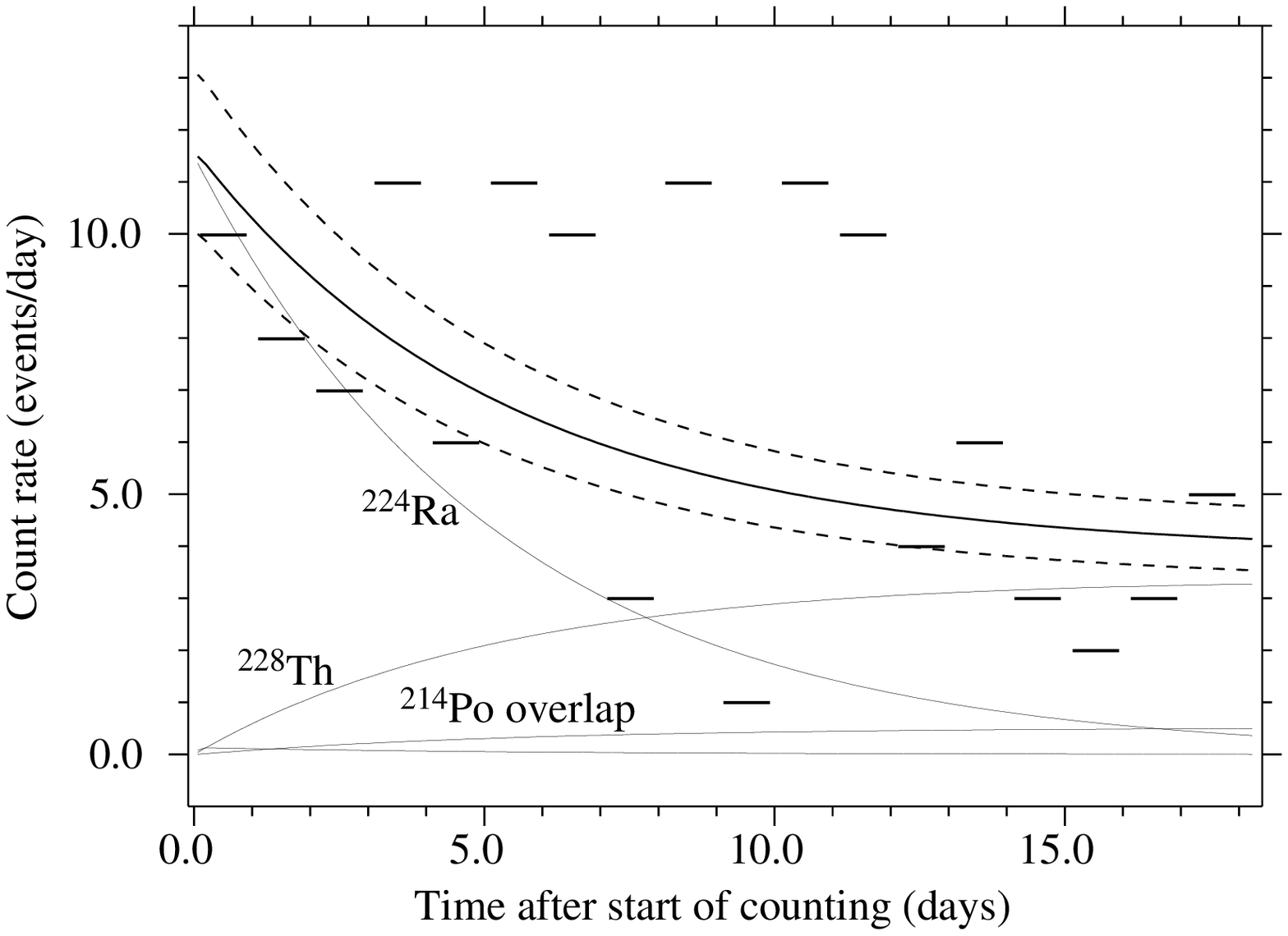}
\includegraphics*[bb=5 26 476 360,width=0.48\hsize]{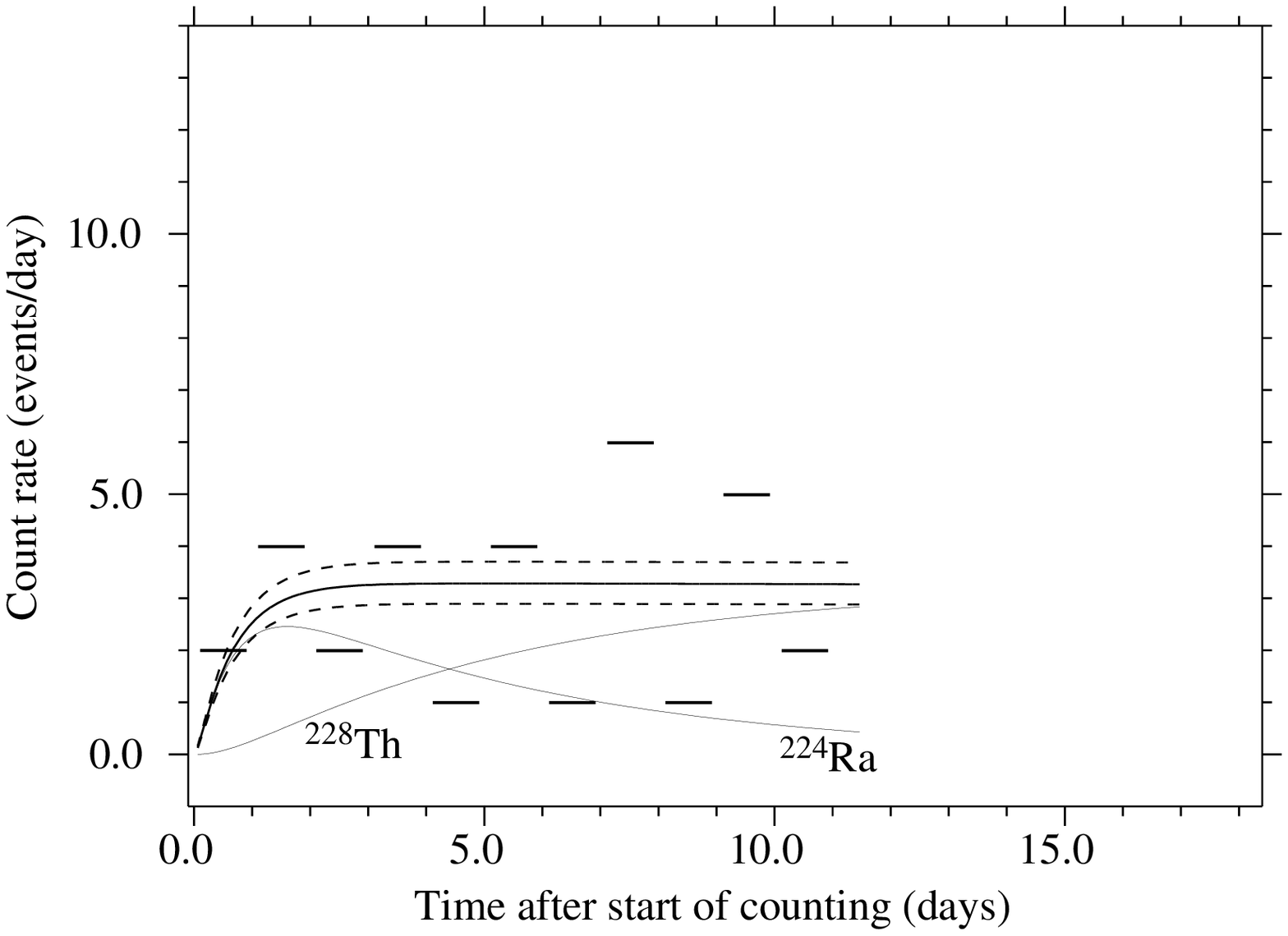}
\hfill%
\includegraphics*[bb=5 26 476 360,width=0.48\hsize]{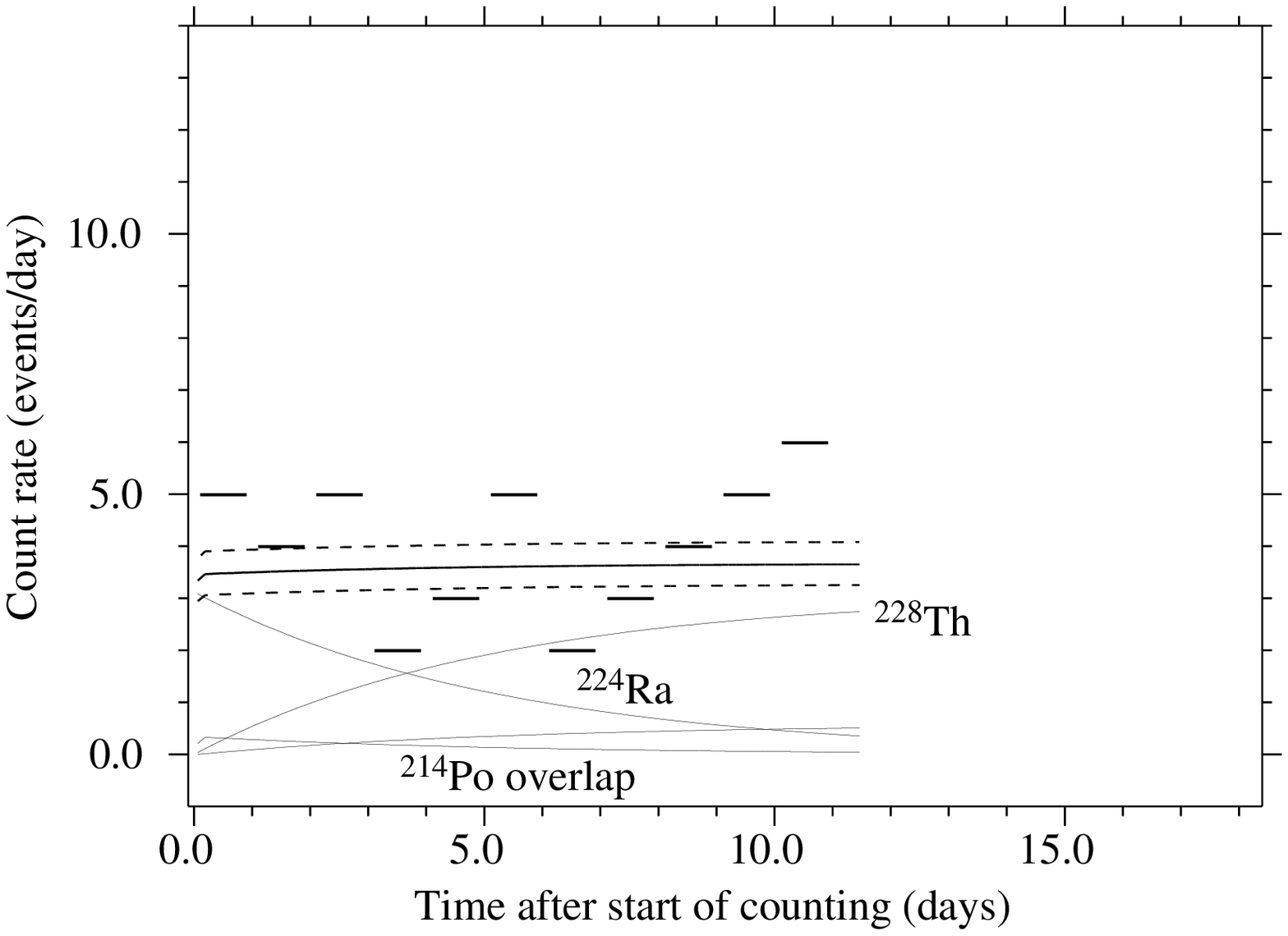}
\caption{Time spectra of \nuc{212}{Po} (left side) and \nuc{216}{Po} (right
side) for the same 4-column \DO assay as in Fig.~\protect\ref{readexample}.
See caption for Fig.~\ref{fitUexample} for further explanation.}
\label{fitThexample}
\end{figure*}

     Examples of time spectra from the last step of energy analysis of the
data in Fig.~\ref{readexample} are shown in Fig.~\ref{fitUexample} for the
\nuc{226}{Ra}-originated components and in Fig.~\ref{fitThexample} for the
\nuc{228}{Th}-originated components.  Fits to these spectra are given by
solid curves with dashed lines above and below the fit that indicate the
range of the fit based on the statistical 68\% confidence limits for the
derived parameters.  The fit to the spectra contains four components for
\nuc{214}{Po} and \nuc{216}{Po}, two components for \nuc{212}{Po}, and six
components for \nuc{218}{Po}.  The major components of the fit are
separately shown by the rising and falling thin solid lines.  The physical
basis for these fits and the time functions used in fitting~\cite{SNO066}
will now be briefly described.

\paragraph{U chain}

     The U chain is the simpler case and will be discussed first.  The
observed Po isotopes, first \nuc{218}{Po} and then \nuc{214}{Po}, are
produced by two sources:
\begin{itemize}
\item[$\bullet$]{The decay of \nuc{222}{Rn} that was entrapped during
      exposure of the column and sections of the ESC counting loop to air.
      Neglecting any permeation of Rn into the loop, the decay rate of this
      component begins at some positive value and falls with the 3.82-day
      half life of \nuc{222}{Rn}.  At time $t$ the instantaneous decay rate
      has the form $a_{222}e^{-\lambda_{222}t}$, where $a_{222}$ is related
      to the initial number of \nuc{222}{Rn} atoms and $\lambda_{222}$ is
      the decay constant of \nuc{222}{Rn}.  This component is shown in the
      spectrum of \nuc{218}{Po} by the thin falling line in
      Fig.~\ref{fitUexample}.  The initial amplitude $a_{222}$ depends on
      the extent of pumpout before the start of counting.  Note that in this
      example there was more trapped Rn in the blank than in the sample.
      The time development of this component is somewhat different in the
      spectrum of \nuc{214}{Po}--there is a delay at the start of counting
      because of the need to build up the daughter elements 26.8-min
      \nuc{214}{Pb} and 19.9-min \nuc{214}{Bi}.  This leads to an initial
      rise in \nuc{214}{Po} count rate before the fall begins.}
\item[$\bullet$]{The decay of \nuc{226}{Ra}.  \nuc{226}{Ra} originates from
      two sources: (1) Ra extracted from the water and (2) contamination
      with Ra of all the components of the ESC loop, viz., the beads, the
      \MnOx coating, the column, the detector walls, etc.  These sources are
      separated by counting the column both before and after extraction and
      taking the difference.  This component has the time dependence in the
      \nuc{218}{Po} spectrum $a_{226}(1 - e^{-\lambda_{222}t})$ where
      $a_{226}$ is related to the initial number of \nuc{226}{Ra} atoms.  It
      is shown by the thin rising lines in Fig.~\ref{fitUexample}.}
\end{itemize}

\paragraph{Th chain}

     The parent isotopes that produce the observed \nuc{216}{Po} and
\nuc{212}{Po} counts are \nuc{228}{Th} and \nuc{224}{Ra}.  These spectra
also originate from two sources:
\begin{itemize}
\item[$\bullet$]{The decay of \nuc{224}{Ra} that (1) was produced by the
      decay of \nuc{228}{Th} and (2) was extracted from the water.  Since it
      is the latter that we wish to determine, the column is counted before
      and after extraction.  In the spectrum of \nuc{216}{Po} this component
      has the time dependence $a_{224}e^{-\lambda_{224}t}$ where $a_{224}$
      is related to the initial number of \nuc{224}{Ra} atoms and
      $\lambda_{224}$ is the \nuc{224}{Ra} decay constant.  It is shown in
      Fig.~\ref{fitThexample} in the spectrum of \nuc{216}{Po} by the thin
      falling line.  In the spectrum of \nuc{212}{Po} it is the component
      that gradually rises to a peak at about 1.5~days after the start of
      counting and then falls with the \nuc{224}{Ra} half life.  The initial
      rise in the \nuc{212}{Po} spectrum is because of the delay in
      production of the 10.6-h daughter \nuc{212}{Pb} from \nuc{224}{Ra}.}
\item[$\bullet$]{The decay of \nuc{228}{Th}.  The \nuc{228}{Th} is present
      as a contaminant in the constituents of the ESC loop and, to a small
      extent, may be extracted from the water.  This term is shown in the
      spectra of Fig.~\ref{fitThexample} by the thin rising line.  In the
      \nuc{216}{Po} spectrum the time dependence is
      $a_{228}(1 - e^{-\lambda_{224}t})$ where $a_{228}$ is related to the
      initial number of \nuc{228}{Th} atoms.  The time dependence in the
      \nuc{212}{Po} spectrum is again delayed by the need to build up
      \nuc{212}{Pb}.}
\end{itemize}

\paragraph{Estimation of activities\newline}
\label{fitsection}

     Both decay chains are examples of the general case $N_1
\xrightarrow[\lambda_1]{} N_2 \xrightarrow[\lambda_2]{} N_3
\xrightarrow[\lambda_3]{} \ldots N_i \xrightarrow[\lambda_i]{} \ldots$ whose
solution is~\cite{BAT10}
\begin{equation}
\label{Bateman}
\lambda_i N_i (t) = N_1(0)\sum_{j=1}^i e^{-\lambda_j t} \lambda_j
              \prod_{\stackrel{\scriptstyle k=1}{\scriptstyle k\ne j}}^i
                                 \frac{\lambda_k}{\lambda_k - \lambda_j},
\end{equation}
where $N_i(t)$ is the number of atoms of species $i$ at time $t$ and the
initial condition $N_i(0)=0$ for all $i\ne1$ is imposed.  The number of
decays $d_i$ of species $i$ during a counting interval that begins at time
$t_{\text{B}}$ and ends at time $t_{\text{E}}$ is thus
\begin{equation}
d_i (t_{\text{B}},t_{\text{E}}) = \int_{t_{\text{B}}}^{t_{\text{E}}}
\lambda_i N_i \diff t
              = N_1(0) \sum_{j=1}^i \Delta_j P_j^i,
\end{equation}
where $\Delta_j = e^{-\lambda_j t_{\text{B}}} - e^{-\lambda_j t_{\text{E}}}$
and $P_j^i$ is the product in Eq.~(\ref{Bateman}).  Introducing the
detection efficiency $\varepsilon_i$ for isotope $i$ and defining the
activity $A_i$ of isotope $i$ to be $A_i = \lambda_i N_i$, the number of
detected counts $m_i$ of species $i$ due to parent isotope 1 during the
counting interval from $t_{\text{B}}$ to $t_{\text{E}}$ is
\begin{equation}
\label{basic}
m_i(t_{\text{B}},t_{\text{E}}) = \varepsilon_i \frac{A_1(0)}{\lambda_1}
\sum_{j=1}^i \Delta_j
               P_j^i.
\end{equation}

     Defining $A_{222}$ as the initial activity of \nuc{222}{Rn} and
$A_{226}$ as the initial activity of \nuc{226}{Ra}, Eq.~(\ref{basic}) is
applied in the U chain to obtain the number of counts of \nuc{218}{Po} and
\nuc{214}{Po} in each counting interval.  Similarly, defining $A_{224}$ as
the initial activity of \nuc{224}{Ra} and $A_{228}$ as the initial activity
of \nuc{228}{Th}, Eq.~(\ref{basic}) is used in the Th chain to obtain the
number of counts of \nuc{216}{Po} and \nuc{212}{Po} in each counting
interval.  Because their half lives are much shorter than the 3-hour
interval on which data is recorded, we make the approximation that the
55.6-sec \nuc{220}{Rn} and 0.15-sec \nuc{216}{Po} decays occur immediately
after the \nuc{224}{Ra} decay.

     In addition to these major terms, there are several additional
components, whose amplitudes are relatively small, which also contribute to
the time spectra: Except for \nuc{212}{Po}, each Po isotope has an overlap
contribution from the adjacent Po isotope at higher energy in the spectrum.
These terms are added onto the expression for
$m_i(t_{\text{B}},t_{\text{E}})$, using for \nuc{216}{Po} the overlap
fraction measured when Rn gas is added to the ESC loop, and for
\nuc{214}{Po} and \nuc{218}{Po} the overlap fraction measured when a Th
source was in the ESC loop.  Finally, for \nuc{218}{Po}, a term is added to
$m_i$ to take account of the contamination of \nuc{212}{Bi}.  The relative
amplitude of this component is approximately determined by the 36\%
branching fraction of \nuc{212}{Bi} to \nuc{208}{Tl} and was checked by fits
to data with a Th source.  All these additional terms are included in the
fitted total spectra shown in Figures~\ref{fitUexample} and
\ref{fitThexample}, but for clarity only the major terms are illustrated.

     For each of the four Po isotopes a likelihood function is formed which
is identical to Eq.~(\ref{LF}) with the factor $m(E_i)$ replaced by the
expression for the total number of predicted counts $m_i$ in
Eq.~(\ref{basic}) and the product taken over all time intervals $i$.  The
grand likelihood function is defined as the product of these individual
likelihood functions for \nuc{212}{Po}, \nuc{214}{Po}, \nuc{216}{Po}, and
\nuc{218}{Po}.  The fit is made by solving for the four variables $A_{222}$,
$A_{224}$, $A_{226}$, and $A_{228}$ that maximize the grand likelihood
function and the parameter confidence regions are found using the same
prescription as described earlier for the energy spectrum.  In the spectra
from a blank the \nuc{224}{Ra} and \nuc{228}{Th} are in equilibrium so the
additional constraint $A_{224} = A_{228}$ is imposed.  In making the fits,
negative values of the parameters are not allowed, since they correspond to
an unphysical regime.

     The final step of time spectrum analysis is to determine the goodness
of fit between the fitted spectra and the data.  For the set of 4 spectra,
the Poisson likelihood $\chi^2$ goodness of fit parameter defined
in~\cite{Baker},
$\chi^2 = 2 \sum_i [m_i - l_i + l_i \ln(l_i/m_i)]$
is calculated, where the sum is over all time intervals $i$ and $l_i$ is the
number of observed counts in interval $i$.  Since there are usually very few
counts in each of the 3-hour data intervals, the probability distribution of
this parameter is not the same as that of the standard $\chi^2$
distribution, but must be obtained by simulation.  This is done by making a
large number of simulations, each of them based on the best fit parameters.
Each simulation is then fit to obtain new parameters and the value of
$\chi^2$ is calculated.  This determines the distribution of $\chi^2$ for
data with the same average number of counts per interval as in the real
data.  The probability level is then set by the placement of the value of
$\chi^2$ from the real data in the distribution from simulation, a procedure
analogous to that described in~\cite{BTC98}.

\begin{table}[ht]
\caption{Data values and fitted parameters for time spectra in
Figs.~\protect\ref{fitUexample} and \protect\ref{fitThexample}.}
\label{fitparameters}
\begin{tabular*}{\hsize}{@{} l @{\extracolsep{\fill}}  r r @{}}
\hline
\hline
                &   \mc{2}{c}{Value}  \\ \cline{2-3}
Item            &   \mc{1}{c}{Blank}   &   \mc{1}{c}{Sample}  \\
\hline
Number of counting intervals &  92  \phantom{xxxi} &  146 \phantom{xxxi} \\
Counting efficiency (\%) \\
\hspace{1em} \nuc{212}{Po}   &  4.4 \phantom{xx}  &  4.4  \phantom{xx} \\
\hspace{1em} \nuc{214}{Po}   & 16.4 \phantom{xx}  & 16.3  \phantom{xx} \\
\hspace{1em} \nuc{216}{Po}   &  4.1 \phantom{xx}  &  4.1  \phantom{xx} \\
\hspace{1em} \nuc{218}{Po}   & 15.1 \phantom{xx}  & 15.1  \phantom{xx} \\
\mc{3}{l}{\hspace{-0.5em} Activity at start of counting (decays/day)}  \\
\hspace{1em} \nuc{224}{Ra}   &  76 $^{+10}_{ -9}$ & 278 $^{+39}_{-37}$ \\
\hspace{1em} \nuc{228}{Th}   &  76 $^{+10}_{ -9}$ &  93 $^{+15}_{-15}$ \\
\hspace{1em} \nuc{226}{Ra}   & 313 $^{+17}_{-17}$ & 280 $^{+10}_{-10}$ \\
\hspace{1em} \nuc{222}{Rn}   & 184 $^{+20}_{-19}$ &  69 $^{+16}_{-15}$ \\
Goodness of fit probability (\%)  & $30 \pm 1.5$  &  $49 \pm 1.6$      \\
\hline
\hline
\end{tabular*}
\end{table}

     The results of fitting the spectra in Fig.~\ref{fitUexample} and
\ref{fitThexample} are given in Table~\ref{fitparameters}.  The values used
for counting efficiency are discussed in Section~\ref{ctgeff}.  There is
considerably more \nuc{224}{Ra} in the sample than in the blank, implying
that \nuc{224}{Ra} was extracted from the water.  Within errors the
\nuc{228}{Th} activity is the same in both sample and blank, indicating that
little Th was extracted.  The disparity in \nuc{222}{Rn} content shows that
there was much less Rn in the column for the sample than in the blank.
Surprisingly, there is also slightly less \nuc{226}{Ra} in the sample than
in the blank, implying that some \nuc{226}{Ra} may have been leached from
the column.  A loss of Ra of this magnitude is consistent with
Eq.~(\ref{Kd}).  The fits to both sample and blank spectra are quite
acceptable, as indicated by the goodness of fit probabilities.

     The results of this example are very much like those of an average
assay.  Typical values for the activities when 4~blank columns are counted
are 75~decays of \nuc{224}{Ra} per day and 250~decays of \nuc{226}{Ra} per
day, with a variation of 20\% and 40\%, respectively.  These should be
compared with measured activities when the ESC loop is counted with no
columns present, but the loop closed with a dummy insert, in which case the
typical activities are 10~decays of \nuc{224}{Ra} per day and
(150--200)~decays of \nuc{226}{Ra} per day.

\subsubsection{Concentration calculation}
\label{conccalc}

     We will now relate the deduced activities of \nuc{224}{Ra} and
\nuc{226}{Ra}, $A_{224}$ and $A_{226}$, respectively, to the concentration
$C_\text{Ra}$ of Ra in the water.  The first step is to determine the net
activity of the sample after subtraction of the blank activity, defined as
\begin{equation}
\label{anet}
A_\text{Ra}^\text{net} = A_\text{Ra}^\text{sample} -
                 A_\text{Ra}^\text{blank}
                 e^{-\lambda_\text{supporting}(t^\text{sample}_\text{SOC} -
                                               t^\text{blank} _\text{SOC})},
\end{equation}
where $A_\text{Ra}^\text{sample}$ is the activity at the start of counting
(SOC) of the sample and $A_\text{Ra}^\text{blank}$ is the activity at the
start of counting of the blank, as calculated by the fit procedure in
Section~\ref{fitsection}.  The exponential factor in this equation decays
the blank activity from the time of counting the blank to the time of
counting the sample, which assumes that the blank activity is not supported
by elements higher in the chain.  This equation applies for both
\nuc{224}{Ra}, in which case $\lambda_\text{supporting}$ is the decay
constant of \nuc{228}{Th}, and \nuc{226}{Ra}, in which case
$\lambda_\text{supporting}$ is the decay constant of \nuc{226}{Ra}.  Since
the time from blank counting to sample counting is usually much less than
the half life of the supporting isotope, the exponential factor is close to
unity.

     A special case arises for \nuc{224}{Ra}: if the \nuc{228}{Th} activity
of the sample exceeds that of the blank, then there is evidence that
\nuc{228}{Th} was extracted from the water.  In this case it is necessary to
subtract from $A_\text{Ra}^\text{net}$ given by Eq.~(\ref{anet}) an
additional term.  See \cite{vers1} for details.

     During the extraction process, the number of atoms of Ra $N$ extracted
from the water and adsorbed on the column is governed by the differential
equation
\begin{equation}
\label{diffeq}
\frac{\diff N}{\diff t} = \varepsilon_\text{extraction}C_\text{Ra} F(t) -
\lambda N(t),
\end{equation}
where $\varepsilon_\text{extraction}$ is the extraction efficiency and
$F(t)$ is the flow rate.  The concentration is assumed here to be a
constant.  Since our assays are made in recirculation, to satisfy this
assumption we must return the water to a point in the vessel that is far
from the sampling point and the volume of water that is sampled must be less
than the total vessel volume.  The general solution of Eq.~(\ref{diffeq})
under the initial condition $N(0)=0$ is
\begin{equation}
\label{DEsolution}
N(t) = \varepsilon_\text{extraction}C_\text{Ra} e^{-\lambda t} \int_0^t
e^{\lambda x} F(x) \diff x.
\end{equation}

     The extraction process is often interrupted and the flow rate may not
be constant.  To model these changes, we break up the extraction time into
$I_\text{extr}$ intervals during each of which the flow rate is constant at
the value $f_i$.  If we define the beginning and ending times of flow
interval $i$ to be $t_{\text{begin}_i}$ and $t_{\text{end}_i}$,
respectively, then the number of Ra atoms that were extracted from the water
and present on the column at the start of counting $t_\text{SOC}$ is
$N(t_\text{SOC}) = \varepsilon_\text{extraction}C_\text{Ra} \Phi$, where
\begin{equation}
\Phi = \frac{1}{\lambda} \sum_{i=1}^{I_{\text{extr}}} f_i
  [ e^{-\lambda(t_\text{SOC} - t_{\text{end}_i})}
   -e^{-\lambda(t_\text{SOC} - t_{\text{begin}_i})} ]
\end{equation}
is called the flow-saturation factor.  $\Phi$ has units of volume.  From the
definition of activity, $A^\text{net} = \lambda_\text{Ra} N(t_\text{SOC})$,
we obtain the Ra concentration
\begin{equation}
C_\text{Ra} = \frac{A_\text{Ra}^\text{net}}
                   {\varepsilon_\text{extraction} \lambda_\text{Ra} \Phi}.
\end{equation}

     As derived here the units of $C_\text{Ra}$ are Ra atoms per unit volume
of water.  It is more customary to express the results in terms of the
parent isotopes at the top of the decay chains, assuming equilibrium.  This
implies in the Th chain $\lambda_{224}C_{224} = \lambda_{232}C_{232}$, where
$C_{232}$ is the concentration of \nuc{232}{Th}, and in the U chain
$\lambda_{226}C_{226} = \lambda_{238}C_{238}$ where $C_{238}$ is the
concentration of \nuc{238}{U}.  Further, we convert to the more common units
of mass per unit volume and remove the isotopic fraction of \nuc{232}{Th}
and \nuc{238}{U}.  This gives the final concentration of Th, $C_\text{Th}$,
and of U, $C_\text{U}$,
\setlength{\arraycolsep}{2pt}
\begin{equation}
\begin{array}{r c l}
C_{\text{Th}} & = & 2.851 \times 10^{-15}
                    \frac{
                          {\displaystyle A_{224}^\text{net}}
                         }
                         {
                          {\displaystyle \varepsilon_\text{extraction}
                                                                 \Phi_{224}}
                         }
                    \frac{
                          {\displaystyle \text{g Th}}
                          }
                          {
                          {\displaystyle \text{cm}^3}
                          }, \\
C_{\text{U}}  & = & 0.937 \times 10^{-15}
                    \frac{
                          {\displaystyle A_{226}^\text{net}}
                         }
                         {
                          {\displaystyle \varepsilon_\text{extraction}
                                                                 \Phi_{226}}
                         }
                    \frac{
                          {\displaystyle \text{g U}}
                         }
                         {
                          {\displaystyle \text{cm}^3}
                         },
\end{array}
\end{equation}
where the activity $A$ is in decays per day and $\Phi$ is measured in
k$\ell$.

\begin{table}[t]
\caption{Factors needed to compute concentration for example data in
Figs.~\protect\ref{fitUexample} and \protect\ref{fitThexample}.}
\label{resultsexample}
\begin{tabular*}{\hsize}{@{} l @{\extracolsep{\fill}} d d @{}}
\hline
\hline
       &   \mc{2}{c}{Value}                             \\ \cline{2-3}
Item   &   \mco{\nuc{224}{Ra}}  &  \mco{\nuc{226}{Ra}}  \rule{0pt}{2.5ex} \\
\hline
Exponential factor in Eq.~({\protect\ref{anet}})      & 0.981 & 1.000     \\
Net activity $A^\text{net}$ (decays/day)              & 197   & <20.0     \\
Flow-saturation factor $\Phi$ (k$\ell$)               & 213.8 & 320.4     \\
Decay constant $\lambda_\text{Ra} (10^{-3} \text{days}^{-1}) $ & 189 &
0.00119\\
Extraction efficiency $\varepsilon_\text{extraction}$ (\%) & 95  & 95     \\
Ra concentration $C_\text{Ra}$ (atoms/k$\ell$)        & 5.1   & <55 200   \\
Parent conc. (10$^{-15}\!$ g Th or U/cm$^3$)          & 2.8   & <0.06     \\
\hline
\hline
\end{tabular*}
\end{table}

     For the example set of data given in Table~\ref{fitparameters}, the
factors needed to compute $C_\text{Ra}$ and the results are listed in
Table~\ref{resultsexample}.  Since the \nuc{226}{Ra} activity of the sample
was less than the blank, we can only derive an upper limit for the
\nuc{226}{Ra} activity, which we set by the quadratic combination of the
errors for the fit of the sample and blank.

\section{Measurement of extraction efficiency}
\label{measureexteff}

     In this section we describe the ways in which the efficiency of
extraction of Ra from water by \MnOx-coated beads has been measured.  The
efficiency is usually very high, 95\% or more.

     The extraction efficiency for Ra is defined as the probability that an
atom of Ra dissolved in the water which flows over the column will be
adsorbed on the column and held there until extraction ends.  As discussed
below in Section~\ref{results}, not all Ra that is present in the system may
be dissolved; further there is a concern that Ra captured on the column at
the start of water flow may be leached from the column by the time flow has
ended, especially in assays that sample a very large volume of water.

     The efficiency of most extractions in the \HO is determined by passing
the water over two columns (or with multicolumn assays, two sets of columns)
in series, called here the `upstream' and `downstream' columns.  Assuming
both columns have the same efficiency, which is valid because we are
considerably  below capacity and the distribution coefficient is independent
of concentration~\cite{ionex}, and no leaching occurs, then it is easily
shown that this efficiency is given by
\begin{equation}
\label{exteff}
\varepsilon_\text{extraction} =
1 - \frac{A_\text{downstream}^\text{net}}{A_\text{upstream}^\text{net}},
\end{equation}
where $A^\text{net}$ is the measured column activity after blank subtraction
and correction for Th extraction.  To help to satisfy the first assumption,
large quantities of the beads, sufficient for filling many columns, are
mixed, and the columns are always filled with the same bead mass.  The
validity of the second assumption, that no leaching occurs, has been tested
by spiking beads with \nuc{226}{Ra} activity and flowing large volumes of
water through them, as described above in Section~\ref{beads}.
Insignificant loss due to leaching was observed up to equivalent volumes of
170~k$\ell$.  Assuming the correctness of these assumptions, we obtain from
Eq.~(\ref{exteff}) in most high-activity assays an efficiency in the range
of (95--100)\%.  In the analysis we use the extraction efficiency from
Eq.~(\ref{exteff}) whenever possible; in other cases an efficiency of 95\%
is assumed with a 5\% systematic uncertainty.

     The extraction efficiency has also been measured by adding weak
\nuc{224}{Ra} spikes to $\sim$1~m$^3$ of water.  By comparison of the known
initial \nuc{224}{Ra} activity with the activity inferred from counting the
column on an ESC, these measurements showed that Ra was routinely extracted
from water by \MnOx with $>$90\% efficiency~\cite{AND95}.

     The extraction efficiency of \MnOx for Th at the flow rate of a
standard assay is quite low ($\aprle$10\%) \cite{Thext}.

\begin{table*}[!bt]
\caption{Factors entering the counting efficiency when ESC is at standard
operating conditions (26~mbar and 1000~V).}
\label{ctgeffact}
\begin{tabular*}{\hsize}{@{} l @{\extracolsep{\fill}} p{0.35\hsize}
p{0.25\hsize} p{0.25\hsize} @{}}
\hline
\hline
       &                       & \mc{2}{c}{Value and uncertainty (\%)}    \\
\cline{3-4}
Factor & \mc{1}{c}{Definition} & \mc{1}{l}{Th chain} & \mc{1}{l}{U chain} \\
\hline
$\varepsilon_\text{emanation}$ &
Probability Rn atom escapes from \MnOx coating and enters gas phase &
$30\pm7$ (\nuc{220}{Rn}) & $75\pm10$ (\nuc{222}{Rn}) \rule{0pt}{2.5ex} \\
$\varepsilon_\text{volume}$ &
Probability Rn atom is in ESC active volume when it decays &
$83.3\pm1.5$ (\nuc{220}{Rn}, 1 column) \newline $55.9\pm1.5$ (\nuc{220}{Rn},
4 columns) &
$81.9\pm1.5$ (\nuc{222}{Rn}, 1 column) \newline $55.4\pm1.5$ (\nuc{222}{Rn},
4 columns) \\
$\varepsilon_\text{detection}$ &
Probability Po atom from Rn decay in ESC active volume produces detected
pulse in $\alpha$ counter &
$26.4\pm3.4$ (\nuc{212}{Po}) \newline $25.6\pm3.3$ (\nuc{216}{Po}) &
$40.7\pm1.8$ (\nuc{214}{Po}) \newline $38.2\pm1.7$ (\nuc{218}{Po}) \\
$\varepsilon_\text{window}$ &
Probability detected $\alpha$ from Po decay is in analysis window for that
Po isotope &
\mc{2}{l}{$97\pm1$ (typical, calculated in each assay for each Po isotope)}
\\
\hline
\hline
\end{tabular*}
\end{table*}

\section{Measurement of counting efficiency}
\label{ctgeff}

     This section gives the results of counting efficiency measurement.  For
short-lived \nuc{220}{Rn}, the major loss of efficiency comes from decay
before it escapes from the column; for long-lived \nuc{222}{Rn}, the major
loss is during the process of detection.  In one-column assays the
approximate counting efficiency is 6\% for the Po isotopes that follow
\nuc{220}{Rn} and 25\% for those that follow \nuc{222}{Rn}.

     The counting efficiency for each Po isotope is defined as the
probability that if an atom of \nuc{226}{Ra} (for \nuc{218}{Po} and
\nuc{214}{Po}) or \nuc{224}{Ra} (for \nuc{216}{Po} and \nuc{212}{Po}) decays
within the column on the ESC, then the resultant Po isotope will yield a
detected count within the energy window chosen for that Po isotope.  This
efficiency was determined in two ways: (1) by \nuc{224}{Ra} spike
experiments similar to those described earlier for measurement of extraction
efficiency and (2) by measurements of all the separate factors that together
make up the counting efficiency.

     The spike technique has the advantage that it measures the entire
throughput of the assay procedure from added Ra to detected Po.  In these
experiments, \nuc{224}{Ra} activity was added to 1~m$^3$ of ultrapure \HO
and the water was flowed through an \MnOx column at 20~$\ell$/min.  The
spike activity was determined by electroplating an aliquot of the same
activity as added to the tank onto a planchet and counting it with an
$\alpha$ detector of known efficiency.  To separately determine the
extraction efficiency, these experiments were made with upstream and
downstream columns that were individually counted.  Removing the extraction
efficiency, which was consistently (98--99)\%, the counting efficiency for
\nuc{216}{Po} was measured~\cite{Carleton} to be $6.0\pm1.7\%$, where the
uncertainty is the standard deviation of the separate measurements.  The
large uncertainty is believed to be mainly due to differences in the \MnOx
coating on the beads, which lead to differences in emanation of Rn.

     During early work to develop the \MnOx method, much attention was
devoted to understanding the various factors that enter the counting
efficiency.  Following the trail from adsorbed Ra to detected Po, this
efficiency can be broken down into the terms defined in
Table~\ref{ctgeffact}.  A brief discussion of these factors and how they are
measured follows.

     The emanation efficiency has been determined by placing beads spiked
with \nuc{224}{Ra} or \nuc{228}{Th} in front of a shielded $\gamma$ counter
and measuring the activity of \nuc{212}{Pb} as Ar or N$_2$ gas was flowed at
low pressure through the beads.  The major purpose of these experiments was
to test Rn emanation with different types of beads and methods of Mn
coating.  The values given in Table~\ref{ctgeffact} are for our present bead
type and operating conditions.  The difference in emanation efficiency
between \nuc{220}{Rn} and \nuc{222}{Rn} is mainly due to the difference in
their half lives.

     The volume efficiency can be calculated based on the known flow rate,
volumes, and pressures in the ESC loop.  At the typical flow rate the
transfer time from the beads to the ESC is (1--2)~sec with 1 column and
(4--10)~sec with 4 columns.  Assuming good mixing, the Rn remains in the ESC
chamber for approximately 25~sec, after which it returns to the column where
it either may be readsorbed on the Mn coating or may make a second pass
through the loop.  Summing the probability of decay in the ESC active volume
for all possible passes leads to the values in Table~\ref{ctgeffact}.
Again, the major cause of efficiency difference between \nuc{220}{Rn} and
\nuc{222}{Rn} is due to their half life difference.  The uncertainty arises
from lack of knowledge of the extent of mixing and the probability of
readsorption, which, from measurements of \nuc{222}{Rn} adsorption in air,
is taken to be~4\% with 1~column.  As a check on these calculations, a
\nuc{224}{Ra}-spiked column was counted by itself and then with~3 other
columns.  The ratio of the count rates in the two configurations was 0.655,
which agrees well with the calculated ratio of $0.671\pm0.022$.

     By adding sources that emit known quantities of \nuc{222}{Rn} or
\nuc{220}{Rn} to an ESC loop, the efficiency product
$\varepsilon_{\text{volume}}\varepsilon_{\text{detection}}\varepsilon_{\text
{window}}$ has been measured~\cite{NIM99} to be $35\pm1.4\%$ for
\nuc{214}{Po} and $22\pm2.8\%$ for \nuc{216}{Po} at our standard operating
conditions.  For the experimental arrangement used in~\cite{NIM99}, the
volume efficiency is calculated to be $90.2\pm1.5\%$ for both \nuc{220}{Rn}
and \nuc{222}{Rn} and the energy window efficiency is calculated to be
95.4\%.  We divide by these two factors and give the resultant detection
efficiency in Table~\ref{ctgeffact}.

     Because of its long half life, essentially all of the charged
\nuc{218}{Po} ions from \nuc{222}{Rn} decay are collected provided the ESC
pressure is $<$100~mbar, as is almost always satisfied during counting.
But, because of the short 150-ms half life of \nuc{216}{Po}, the daughter of
\nuc{220}{Rn}, the efficiency for \nuc{216}{Po} detection depends more
sensitively on the N$_2$ pressure in the counting loop.  Measured values are
given in \cite{vers1}.  The pressure in the ESC counting loop generally
rises less than 1~mbar/day which leads to a reduction in the \nuc{216}{Po}
detection efficiency of no more than a few percent during the course of a
normal 10-day counting period.  One might expect that the detection
efficiency for \nuc{212}{Po} would be approximately 64\% of that for
\nuc{216}{Po}, because of the 64\% branching ratio of \nuc{212}{Bi} to
\nuc{212}{Po}.  Measurements, however, show that the \nuc{212}{Po}
efficiency is somewhat greater than that of \nuc{216}{Po}, presumably due to
collection of some of the \nuc{212}{Pb} ions if the \nuc{216}{Po} decays in
flight before it reaches the diode (and perhaps to collection of some
charged \nuc{212}{Bi} atoms which recoil into the gas when \nuc{212}{Pb} on
a wall decays).  Measurements with a Th foil in the ESC loop gave a
\nuc{212}{Po} to \nuc{216}{Po} efficiency ratio of 1.034.  For similar
reasons, the \nuc{214}{Po} efficiency is somewhat higher than the
\nuc{218}{Po} efficiency.  Measurements with \nuc{222}{Rn} in six different
ESCs gave a \nuc{214}{Po} to \nuc{218}{Po} efficiency ratio of
$1.065\pm0.006$.

     The final efficiency factor, the window efficiency, is calculated by
numerical integration over the chosen energy window using the fitted line
shape parameters.  Because the energy windows are set as wide as possible,
its value is quite high, typically 98\% for \nuc{212}{Po}, 97\% for
\nuc{214}{Po} and (95--96)\% for \nuc{216}{Po} and \nuc{218}{Po}.

     By multiplying the first three factors listed in Table~\ref{ctgeffact},
this second method of efficiency determination gives a counting efficiency
in 1-column assays of $6.4\pm1.7\%$ for \nuc{216}{Po} and $25.0\pm3.5\%$ for
\nuc{214}{Po}.  The window efficiency is excluded here because it is
calculated separately for each experiment and each energy window.  These are
the standard counting efficiencies that we use for data analysis.  Including
a typical 95\% energy window efficiency, the \nuc{216}{Po} efficiency is
$6.1\pm1.6\%$ which agrees well with the efficiency measured by adding
\nuc{224}{Ra} spikes to water of $6.0\pm1.7\%$.  In 4-column assays the
counting efficiency is reduced by factors of 0.67 for \nuc{216}{Po} and 0.68
for \nuc{214}{Po}.

\section{Systematic uncertainties}

     The various systematic effects are listed in
Table~\ref{allsystematics}.  The largest entry is the 27\% uncertainty in
the counting efficiency for \nuc{224}{Ra}, discussed in Sec.~\ref{ctgeff}.
The other major entries are briefly considered here; the terms whose
uncertainties are 5\% or less are described in \cite{vers1}.

\begin{table}[ht]
\caption{Sources of systematic uncertainty in Ra concentration.  All entries
are symmetric unless otherwise indicated.}
\label{allsystematics}
\begin{tabular*}{\hsize}{@{} l @{\extracolsep{\fill}} @{\hspace{-0.25em}} r
r r r}
\hline
\hline
                    & \mc{4}{c}{Syst.\ uncertainty (\%)} \\ \cline{2-5}
\rule{0cm}{2.5ex}   & \mc{2}{c}{\nuc{224}{Ra}} & \mc{2}{c}{\nuc{226}{Ra}} \\
\cline{2-3} \cline{4-5}
Contributor                  &  \DO &  \HO &  \DO &  \HO  \\
\hline
Water flow \\
     \hspace{0.5em}
  Extraction efficiency      &     5 &     5 &     5 &     5  \\
     \hspace{0.5em}
  Water sample volume        &     5 &     2 &     5 &     2  \\
     \hspace{0.5em}
  Resampling correction      & $<$+4 &     0 & $<$+4 &     0  \\
     \hspace{0.5em}
  Feedback correction        &     3 &     3 &     3 &     3  \\
     \hspace{0.5em}
  \MnOx fines recirculation  &    +5 &    +5 &    +5 &    +5  \\
     \hspace{0.5em}
  Ra leaching                &   +10 &    +3 &   +10 &    +3  \\
     \hspace{0.5em}
  Ra adsorption              & $<$+2 & $<$+2 & $<$+2 & $<$+2  \\
     \hspace{0.5em}
  Ra background (see text) \\
Data acquisition \\
     \hspace{0.5em}
  Counting efficiency        &    27 &    27 &    14 &    14  \\
     \hspace{0.5em}
  Relative ESC counting eff. &     5 &     5 &     5 &     5  \\
     \hspace{0.5em}
  Ra distribution in column  &    +5 &    +5 &    +1 &    +1  \\
     \hspace{0.5em}
  Drifts during counting     &     2 &     2 &     1 &     1  \\
Data analysis \\
     \hspace{0.5em}
  Overlap                    & $<$ 1 & $<$ 1 & $<$ 1 & $<$ 1  \\
     \hspace{0.5em}
  Relative line efficiencies &     3 &     3 &     0 &     0  \\
     \hspace{0.5em}
  Time functions             &     5 &     5 &     5 &     5  \\
\hline
Total,                       &$<$+32 &$<$+30 &$<$+21 &$<$+18  \\
     \hspace{0.5em}
combined quadratically       &   -29 &   -29 &   -17 &   -17  \\
\hline
\hline
\end{tabular*}
\end{table}

     In some high-volume assays a lower \nuc{226}{Ra} activity is seen in
the sample than in the blank, implying that some leaching may be occurring.
If this is true, then \nuc{224}{Ra} must of course also be lost, but no
decrease in \nuc{224}{Ra} activity is seen.  This is because the ratio of
concentrations between water and column is higher for \nuc{224}{Ra} than for
\nuc{226}{Ra}, so the loss of \nuc{224}{Ra} due to leaching is less than the
pickup from the water.  Measurements very rarely show more than a 10\% loss
of \nuc{226}{Ra} activity, so we assign a +10\% uncertainty for Ra leaching
in \DO extractions.  This effect is smaller in \HO extractions as much less
water is circulated.

     The \nuc{224}{Ra} background from \nuc{228}{Th} plated on the walls of
the extraction apparatus of the \DO system was measured as follows: A closed
loop was first made whose water flow bypassed the acrylic vessel.  The \DO
in this loop was flowed through an \MnOx column to extract all Ra in the
water, and the loop was left without flow for 4~days, the typical assay
duration.  Then any fresh Ra that had grown in was extracted by circulating
the water in the loop over an \MnOx column for 3~hours.  After blank
subtraction the net activity was measured to be $16^{+28}_{-16}$~decays of
\nuc{224}{Ra}/day.  This should be compared with the activity extracted in a
typical assay, which is $200 \pm 45$~decays of \nuc{224}{Ra}/day.  The
background from this source is thus $8^{+14}_{\phantom{1}-8}\%$ of the
typical assay activity.  Adopting the upper limit as the uncertainty and
converting to the equivalent concentration of \nuc{232}{Th}, we find that
this systematic background is -0.6\E{-15}~g~Th/cm$^3$~\DO.

\section{Results of water assays}
\label{results}

     We give here the results of \MnOx assays of the \nuc{224}{Ra} and
\nuc{226}{Ra} concentrations in the heavy and light water of the \SNO
experiment.

\renewcommand{\E}[1]{$\times10^{#1}$\xspace}
\begin{table}[ht]
\caption{Equivalent Th and U content for a few typical 4-column \MnOx assays
of \DO inside the acrylic vessel.  Errors are statistical with 68\%
confidence.  The systematic error is given in the 2nd and 4th columns of
Table~\ref{allsystematics}.  To convert from the units used here,
g~impurity/cm$^3$ \DO, to units of g~impurity/g \DO, divide by the \DO
density of 1.106~g/cm$^3$ at 13\degreesC, the temperature of the \DO when it
flows through the columns.}
\label{d2oresults}
\begin{tabular*}{\hsize}{@{} l @{\extracolsep{\fill}} d @{\hspace{-0.5em}} c
c @{}}
\hline
\hline
Assay        &  \mco{Volume}                                         \\
Date         &  \mco{\DO}                                            \\
(yy/mm/dd)   &  \mco{(k$\ell$)}    & g Th/cm$^3$ \DO  & g U/cm$^3$ \DO    \\
\hline
00/05/01 & 241.5 &
        4.9\E{-15}(1$^{+0.27}_{-0.38}$) & 4.1\E{-15}(1$^{+0.09}_{-0.09}$) \\
00/08/22 & 297.0 &
        2.3\E{-15}(1$^{+0.49}_{-0.57}$) & 3.9\E{-16}(1$^{+0.56}_{-0.55}$) \\
00/10/30 & 319.4 &
        2.5\E{-15}(1$^{+0.26}_{-0.27}$) & 0.0(0.0--9.6\E{-17})            \\
01/01/30 & 381.3 &
        2.8\E{-15}(1$^{+0.20}_{-0.21}$) & 0.0(0.0--6.2\E{-17})            \\
01/04/16 & 383.6 &
        2.1\E{-15}(1$^{+0.23}_{-0.23}$) & 1.9\E{-16}(1$^{+0.20}_{-0.20}$) \\
01/05/14 & 434.1 &
        2.2\E{-15}(1$^{+0.18}_{-0.17}$) & 0.0(0.0--4.0\E{-17})            \\
\hline
\hline
\end{tabular*}
\end{table}

     The results of several 4-column assays of the \DO are given in
Table~\ref{d2oresults}.  To infer the average concentration of Th and U in
the heavy water from these results requires an understanding of flow
patterns, plating, solubility, diffusion, etc. and is the subject of a
future article.  Nonetheless, if one compares these results to the upper
limit goals for radioactive purity given in the Introduction, it is apparent
that the U value is considerably below the goal and the Th measurement is
nearly a factor of 2 less than the goal.

\begin{figure}[t]
\includegraphics*[bb=22 72 476 368,width=\hsize]{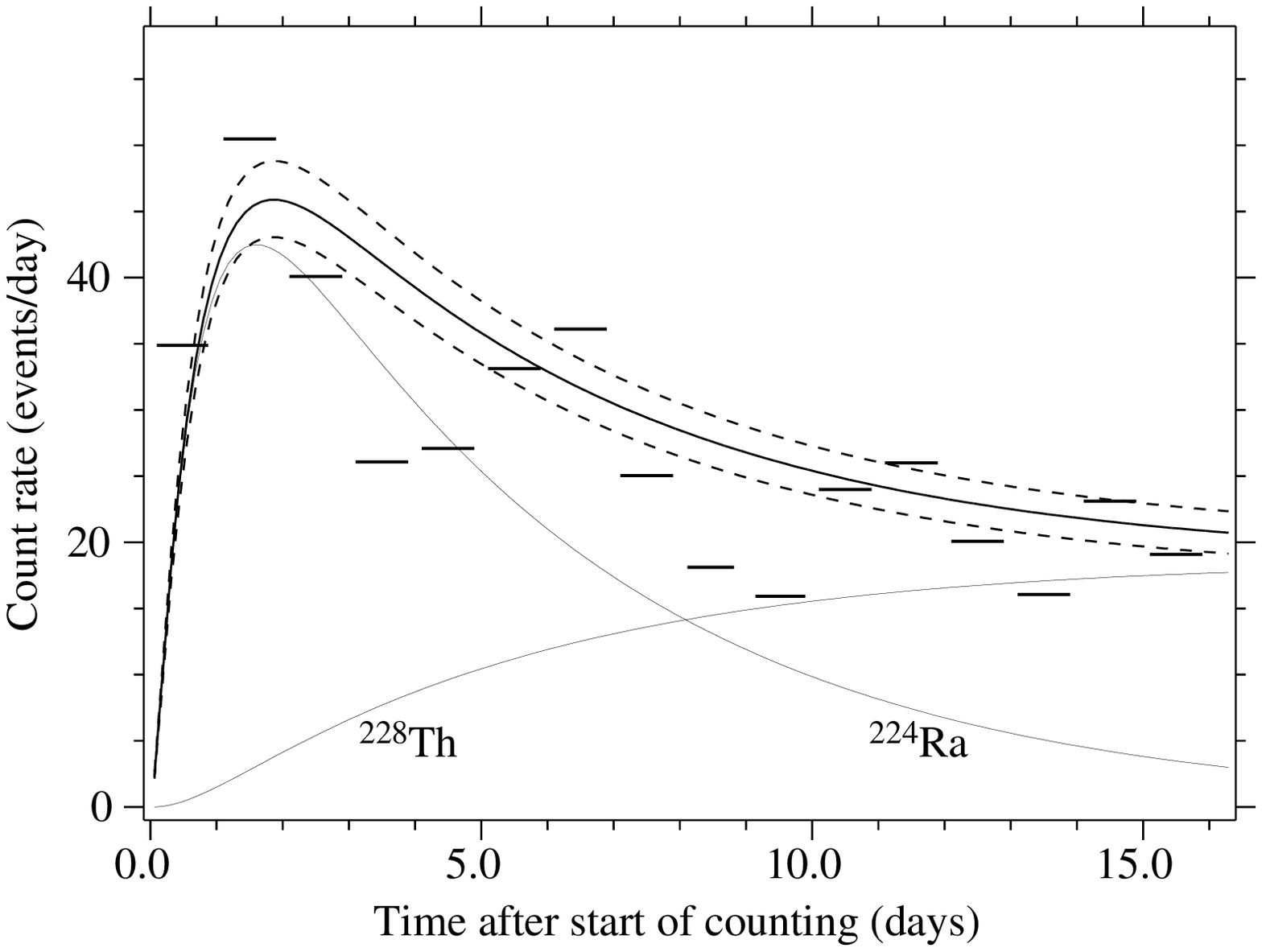}
\includegraphics*[bb=22 26 476 360,width=\hsize]{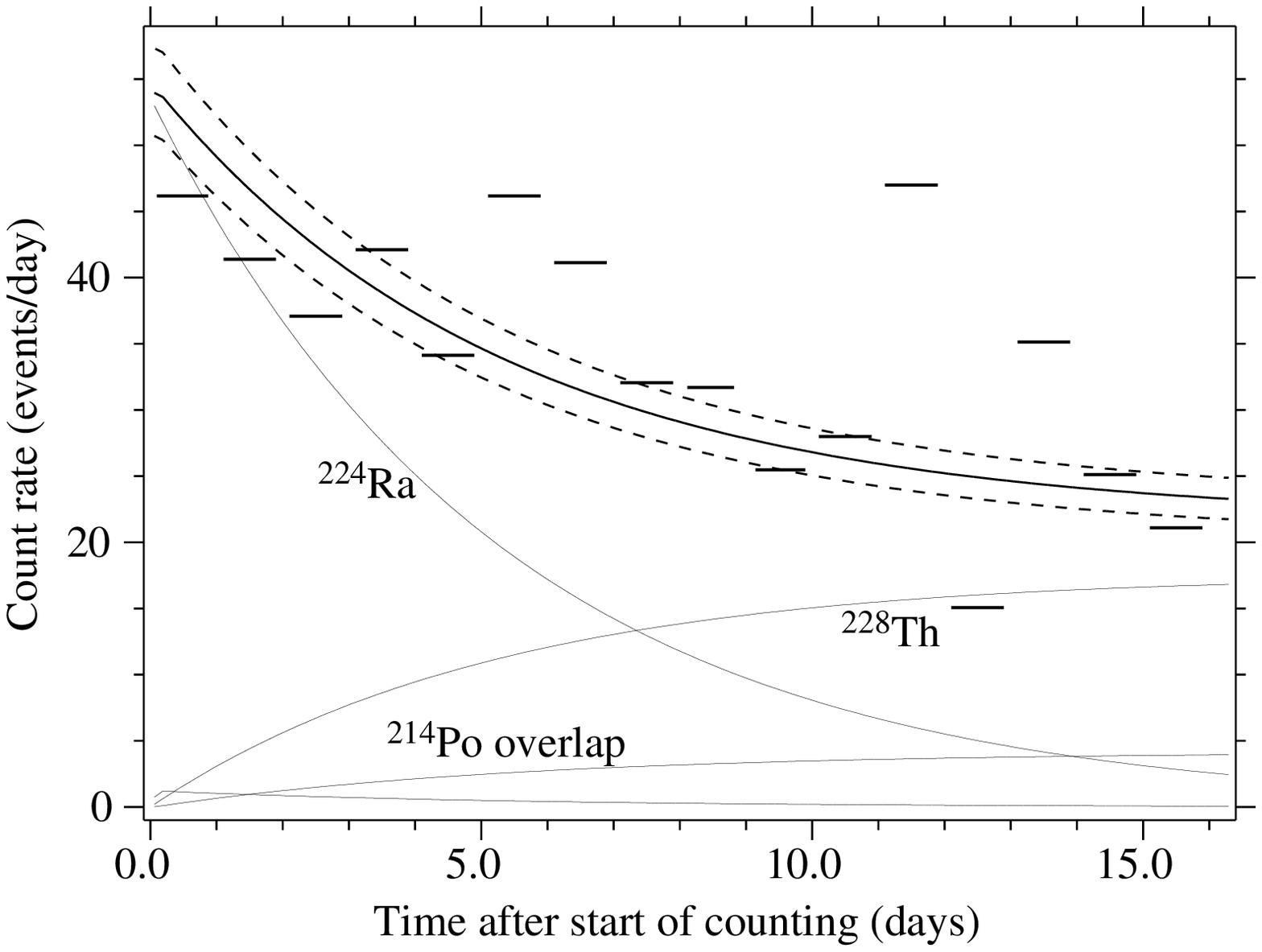}
\caption{Combined time spectra and fit for \nuc{212}{Po} (top) and
\nuc{216}{Po} (bottom) for six 4-column \DO assays.  The data from every
eight 3-hour data collection intervals have been combined.}
\label{sumsum}
\end{figure}

     A useful way to visualize the signal of extracted \nuc{224}{Ra} is to
combine the data from several separate measurements.  This can be done by
simply adding the counts in the separate time intervals after start of
counting for each experiment, with appropriate corrections for times when
the counter was not active.  Figure~\ref{sumsum} shows the \nuc{212}{Po} and
\nuc{216}{Po} spectra obtained by combining in this manner six 4-column
assays of \DO.  The decay of \nuc{224}{Ra} is much more apparent than for
the similar plot for a single measurement in Fig.~\ref{fitThexample}.

\begin{table}[ht]
\caption{Equivalent Th and U content for a few typical 1-column \MnOx assays
of \HO.  Errors are statistical with 68\% confidence.  The systematic error
is given in the 3rd and 5th columns of Table~\ref{allsystematics}.}
\label{h2oresults}
\begin{tabular*}{\hsize}{@{} l @{\extracolsep{\fill}} d @{\hspace{-0.5em}} c
c @{}}
\hline
\hline
Assay        &  \mco{Volume}                                         \\
Date         &  \mco{\HO}                                            \\
(yy/mm/dd)   &  \mco{(k$\ell$)}    & g Th/cm$^3$ \HO  & g U/cm$^3$ \HO    \\
\hline
\hline
\mc{4}{c}{Water between acrylic vessel and photomultipliers} \\
\hline
\mc{4}{l}{Upstream column} \\
\hline
00/01/18 &  51.5 &
        1.0\E{-13}(1$^{+0.14}_{-0.14}$) & 4.9\E{-15}(1$^{+0.09}_{-0.09}$) \\
00/05/10 &  47.0 &
        1.3\E{-13}(1$^{+0.07}_{-0.07}$) & 2.8\E{-15}(1$^{+0.22}_{-0.22}$) \\
00/10/23 & 110.2 &
        6.7\E{-14}(1$^{+0.06}_{-0.06}$) & 1.1\E{-15}(1$^{+0.14}_{-0.14}$) \\
01/02/05 &  45.6 &
        5.9\E{-14}(1$^{+0.08}_{-0.08}$) & 0.0(0.0--2.4\E{-16})            \\
01/05/23 &  51.9 &
        6.1\E{-14}(1$^{+0.08}_{-0.08}$) & 1.3\E{-15}(1$^{+0.15}_{-0.14}$) \\
\hline
\mc{4}{l}{Downstream column} \\
\hline
00/01/18 &  51.5 &
        3.6\E{-15}(1$^{+0.54}_{-0.48}$) & 0.0(0.0--4.2\E{-16})            \\
00/05/10 &  47.0 &
        2.0\E{-15}(1$^{+0.89}_{-0.68}$) & 0.0(0.0--5.6\E{-16})            \\
00/10/23 & 110.2 &
        2.3\E{-15}(1$^{+0.50}_{-0.44}$) & 0.0(0.0--1.4\E{-16})            \\
01/02/05 &  45.6 &
        7.7\E{-16}(1$^{+2.67}_{-1.00}$) & 0.0(0.0--2.7\E{-16})            \\
01/05/23 &  51.9 &
        0.0(0.0--1.2\E{-15})            & 0.0(0.0--1.9\E{-16})            \\
\hline
\hline
\mc{4}{c}{Water outside photomultipliers} \\
\hline
\mc{4}{l}{Upstream column} \\
\hline
99/05/10 &  55.5 &
        1.7\E{-13}(1$^{+0.10}_{-0.09}$) & 1.5\E{-14}(1$^{+0.06}_{-0.06}$) \\
00/05/15 &  34.9 &
        1.2\E{-13}(1$^{+0.09}_{-0.08}$) & 0.0(0.0--6.2\E{-16})            \\
01/04/05 &  14.2 &
        8.4\E{-14}(1$^{+0.18}_{-0.17}$) & 3.1\E{-15}(1$^{+0.39}_{-0.38}$) \\
\hline
\mc{4}{l}{Downstream column} \\
\hline
99/05/10 &  55.5 &
        4.6\E{-14}(1$^{+0.22}_{-0.20}$) & 4.2\E{-15}(1$^{+0.19}_{-0.19}$) \\
00/05/15 &  34.9 &
        5.8\E{-15}(1$^{+0.59}_{-0.51}$) & 0.0(0.0--6.1\E{-16})            \\
01/04/05 &  14.2 &
        1.3\E{-14}(1$^{+0.68}_{-0.61}$) & 1.9\E{-14}(1$^{+0.09}_{-0.09}$) \\
\hline
\hline
\end{tabular*}
\end{table}
\renewcommand{\E}[1]{$\,\times\,10^{#1}$\xspace}

     Two types of Ra assays are often made in the \HO system.  The first has
a sampling point on the midplane of the acrylic vessel halfway between the
vessel and the photomultipliers with the purified water returned to the
midplane at two points, with the closest at 7.0~m distance from where the
sample is taken.  Since the Ra concentration in this region is higher than
in the \DO, an adequate \nuc{224}{Ra} signal is obtained with a single \MnOx
column and a sample volume of only $\sim$50~k$\ell$.  The results of several
measurements of this type are given in Table~\ref{h2oresults}.  Both
upstream and downstream columns are usually used; applying
Eq.~(\ref{exteff}) to \nuc{224}{Ra}, it is seen that the extraction
efficiency is in the range of (96--98)\%.  The other type of frequent \HO
assay is of the water in the region outside the photomultipliers.  The
sampling point is 4~m below the surface and the water is returned to either
the midplane of the acrylic vessel or near the water surface.  Some
representative results are given in Table~\ref{h2oresults}.  Comparing the
results for the region between the acrylic vessel and the photomultipliers
to the upper limit goals for radioactive purity stated in the Introduction,
one sees that the U content of the bulk \HO amply meets the goal and the Th
content is slightly, but not significantly, higher than the goal.

     For all the \DO assays in Table~\ref{d2oresults}, and all but one of
the \HO assays in Table~\ref{h2oresults}, the \nuc{228}{Th} activity of
sample and blank were equal within the statistical 68\% confidence range.
This is expected as the extraction efficiency for Th is low, and most Th, if
present, is expected to be plated onto surfaces, or possibly be held in the
water in the form of a soluble chemical complex.

\section{Conclusion}

     The \MnOx assay method has shown that the Ra activity in the bulk \DO
and the bulk \HO of the \SNO detector is sufficiently low that the rate of
photodisintegration background events produced by Ra impurities should be no
more than 10\% of the rate of neutral-current solar neutrino events
predicted by the standard solar model.  The sensitivity of the \MnOx assay
method is thus adequate for the needs of the \SNO detector.

     Since the \MnOx method is reliable and robust, \SNO has used it to make
many water quality measurements in addition to those in Sec.~\ref{results}.
These experiments mostly tested individual sections of the water system,
such as the piping from the acrylic vessel to the \MnOx columns, the
concentrate of the reverse osmosis apparatus, the contents of various
storage tanks, the purity of the \DO during detector filling, etc.  A
desirable feature of the method is that after extraction the column is
simply dried and counted; there is no need for any chemical processing,
which may introduce background from reagents and the processing system.

     Defining the detection limit as $\sim$3 times the standard deviation of
the background~\cite{Currie}, then, as can be seen from the results in
Table~\ref{fitparameters} for a typical assay, the ultimate sensitivity of
the \MnOx method in a single measurement, as presently used, is
$\sim$5\E{-16}~g~Th/cm$^3$ \DO and $\sim$2\E{-16}~g~U/cm$^3$ \DO.

\section*{Acknowledgments}

     We thank the operators of the \SNO water systems for their great care
in carrying out the assays described here.  We are very grateful to the
INCO, Ltd.\ mining company and their staff at the Creighton mine without
whose help this work could not have been conducted.  We greatly thank Atomic
Energy of Canada, Ltd.\ for the loan of the heavy water in cooperation with
Ontario Power Generation.  This research was supported by: Canada: NSERC,
NRC, Industry Canada, the Northern Ontario Heritage Fund Corporation, and
INCO; USA: Dept.\ of Energy; and UK: PPARC.

\end{document}